\documentclass{CSML}

\pdfoutput=1

\usepackage{lastpage}
\def\dOi{13(4:8)2017}
\lmcsheading%
{\dOi}
{1--\pageref{LastPage}}
{}
{}
{Jul.~27, 2016}
{Nov.~15, 2017}
{}

\usepackage{amsmath}
\usepackage{amssymb}
\usepackage{meq}
\usepackage{xspace}

\usepackage[hidelinks]{hyperref}

\usepackage[pdftex]{color}
\definecolor{mycolor}{gray}{0.8}
\usepackage{header}
\usepackage{framed}

\usepackage[all]{xy}
\usepackage[UglyObsolete]{diagrams}
\diagramstyle[repositionpullbacks,midshaft,scriptlabels]
\newarrowhead{mapsto}\rtbar\ltbar\dtbar\utbar
\newarrow{HH}{mapsto}---{mapsto}

\usepackage{fancyvrb}
\usepackage{here}

\usepackage{wrapfloat}
\usepackage{url}

\setlist{itemsep=.5em,topsep=.5em}
\begin{document}
 
\title[Cyclic Datatypes modulo Bisimulation]{%
Cyclic Datatypes modulo Bisimulation based on Second-Order Algebraic Theories
}

\author[M.~Hamana]{Makoto Hamana}
\address{Department of Computer Science, Gunma University}

\email{hamana@cs.gunma-u.ac.jp}

\maketitle

\begin{abstract} 

Cyclic data structures, such as cyclic lists, in functional programming
are tricky to handle because of their cyclicity. This paper presents an
investigation of categorical, algebraic, and computational foundations of
cyclic datatypes. Our framework of cyclic datatypes is based on
second-order algebraic theories of Fiore et al., which give a uniform
setting for syntax, types, and computation rules for describing and
reasoning about cyclic datatypes.  We extract the ``fold'' computation
rules from the categorical semantics based on iteration categories of
Bloom and \Esik. Thereby, the rules are correct by construction. We prove
strong normalisation using the General Schema criterion for second-order
computation rules.  Rather than the fixed point law, we particularly
choose \Bekic law for computation, which is a key to obtaining strong
normalisation.  We also prove the property of ``Church-Rosser modulo
bisimulation'' for the computation rules.  Combining these results, we
have a remarkable decidability result of the equational theory of
cyclic data and fold.

\end{abstract}

% Local Variables:
% TeX-master: "fs"
% End:

\section{Introduction}\label{sec:intro}
Cyclic data structures in functional programming
are tricky to handle. %because of their cyclicity. 
In Haskell, one can define a cyclic data structure, such as 
cyclic lists by 
\begin{verbatim}
      clist = 2:1:clist
\end{verbatim}
The feasibility of
such a recursive definition of cyclic data
depends on lazy evaluation. 
For example, one can safely take the head of the cyclic list:
\begin{Verbatim}[commandchars=\\\{\},codes=\mathcom]
      head clist $\quad\narone\quad$ 2
\end{Verbatim}
However, this encoding is \W{not completely safe}.
For example, consider the sum of all elements 
using the above 
recursive encoding.
of \code{clist} in Haskell. It falls into non-termination:
\begin{Verbatim}[commandchars=\\\{\},codes=\mathcom]
      sum clist $\quad\narone\quad$ \W{non-termination}
\end{Verbatim}
This means that such a naive encoding of cyclic structure does not 
ensure safety.

The computation using our framework is guaranteed to be \W{safe} meaning that
it is always
terminating, i.e., \W{strongly normalising}.
We provide a way to regard the sum of a cyclic list 
as a {cyclic natural number}, which is computed by the strongly
normalising ``fold'' combinator.
In this paper, we develop a framework for
syntax and semantics of \W{cyclic datatypes} that makes this understanding
and computation correct.

Our framework of cyclic datatypes is founded on  second-order
algebraic theories of Fiore et al. \cite{2ndCSL,2ndAlg}.
Second-order algebraic theories are founded on 
the mathematical theory of second-order 
abstract syntax by Fiore, Plotkin, and Turi
\cite{FPT,free,Fiore2nd} and have been shown to be a useful framework
that models various important notions of programming languages, such as
logic programming \cite{SamFOS}, algebraic effects \cite{SamLICS}, 
quantum computation \cite{SamQ}.
This paper gives another application of second-order algebraic theories,
namely, to cyclic datatypes and their computation.
We use second-order algebraic theories to 
give a uniform setting for typed syntax, equational logic and
computation rules for describing and reasoning about cyclic datatypes.
We extract computation rules for the fold from the categorical semantics
based on iteration categories %of Bloom and Esik 
\cite{BE}. 
Thereby the rules are
correct by construction. Finally, we prove strong normalisation by
using the General Schema criterion \cite{IDTS00} for second-order
computation rules.

\begin{rulefigt}
\begin{center}
\includegraphics[scale=.46]{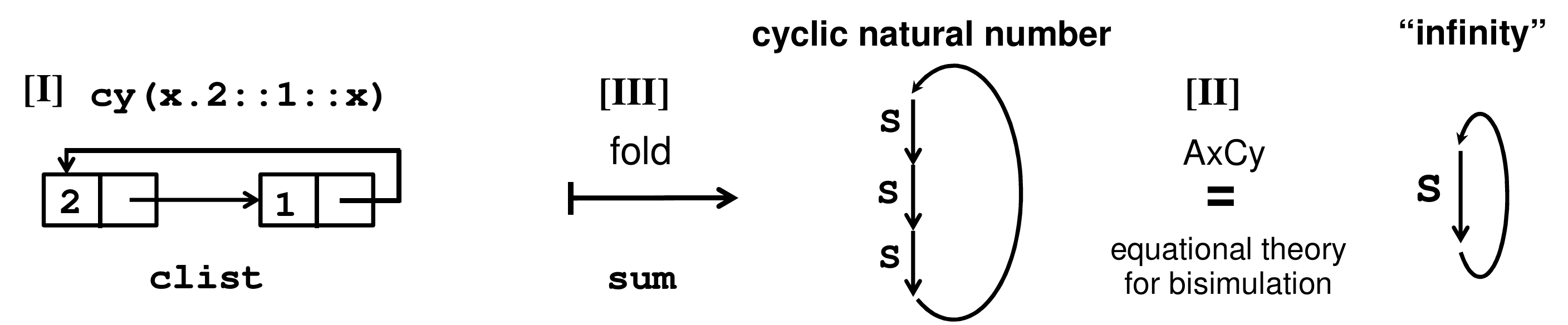}
\end{center}
\caption{Framework: Second-order algebraic theories and iteration theories}
\label{fig:clist-view}
\end{rulefigt}

\subsec{Overview of computation}
As an overview of cyclic datatypes and their operations we develop
in this paper, we first demonstrate descriptions and an operation of
cyclic datatypes by  pseudo-program codes.
The code fragments 
correspond one-to-one to theoretical data given in later sections.
Therefore, they are theoretically meaningful, 
while intuitively understandable without going into the
detailed theory.

First we consider an example of cyclic lists.
The codes below with the keyword \code{ctype}
are intended to declare cyclic datatypes. 
Here we declare the type \code{CNat} of natural numbers
and the type \code{CList} of cyclic lists 
having ordinary constructors in Haskell or Agda style.
\medskip
\begin{center}
\begin{minipage}{0.4\textwidth}
\begin{Verbatim}[commandchars=\\\{\},codes=\mathcom]
ctype CNat where
  0 : CNat
  S : CNat \tto CNat
with axioms \AxCy
\end{Verbatim}
\end{minipage}
\begin{minipage}{0.4\textwidth}
\begin{CVerbatim}[commandchars=\\\{\},codes=\mathcom]
ctype \List where
  $[\, ]$ : \List
  $::$  : CNat,$\List \to \List$
with axioms \AxG
\end{CVerbatim}
\end{minipage}
\end{center}
\medskip
We assume that any \code{ctype} declared datatype has a default
constructor ``\code{cy}'' for making a cycle. For example,
we express a cyclic list of $1$ as a term
$$\code{cy(x.\cons 1 x)}$$
where \code{cy} has 
a variable binding ``\code{x.}'', regarded as 
the ``address'' of the top of list.
This is a fundamental idea presented in \cite{DBLP:conf/popl/FegarasS96}
\cite{TFPcyc}.

A variable occurrence \code x in the body refers to the top,
hence it makes a cycle. 
The terms built from the constructors of \code{CList} and 
the default constructor \code{cy} are required to
satisfy the axioms \AxG (given later in Fig. \ref{fig:axioms})
indicated by the keyword ``\code{with axioms}'' (we assume that any 
\code{ctype} datatype satisfies \AxG, so this is for ease of understanding).
We next consider the above mentioned example of the sum of 
cyclic list.

\begin{CVerbatim}[commandchars=\\\{\},codes=\mathcom]
sum : \List \tto CNat
spec sum (\nil)   = 0
     sum (k \con t) = plus(k, sum (t))
\end{CVerbatim}

\noindent
The above code with the keyword \code{spec}
describes an equational \W{specification} 
of a function.
It requires that the sum function from cyclic lists to cyclic natural numbers
must satisfy the usual recursive properties of sum.
We intend that the \code{spec} code is
merely a (loose) specification, and not a definition,
because it lacks the case of a \code{cy}-term.
Here we assume that the \code{plus} function on \code{CNat} has already 
been defined
(as presented later in Example \ref{ex:plus}).
The following code with the keyword \code{fun} defines
the function \code{sum}. 

\begin{CVerbatim}[commandchars=\\\{\},codes=\mathcom]
  fun sum t = fold (0, k.x.plus(k,x)) t
\end{CVerbatim}

\noindent
It is defined by the \code{fold} combinator on the cyclic datatypes.
The first two arguments \code 0 and \code{k.x.plus(k,x)} correspond to
the right-hand sides of the specification of \code{sum}, where
\code{k.x.} are variable binders (as in \lmd-terms).
The \code{fold} is actually the fold on a cyclic datatype, which knows
how to cope with \code{cy}-terms.
The sum of a cyclic list can be computed as follows:

\begin{CVerbatim}[commandchars=\\\{\},codes=\mathcom]
  sum(cy(x.S$^2$(0) \con S(0) \con x)) 
  \ttoo cy(x. fold (0, k.x.plus(k,x)) (x.S$^2$(0) \con S(0) \con x;x)) 
  \ttoo cy(x.S(S(S(x))))
\end{CVerbatim}

\noindent
where we represent a usual natural number $n$ by \code{S$^n$(0)}.
The final term is a normal form that cannot be rewritten further.
Therefore, we regard it as the computation result.
The steps presented above are actual rewrite steps by the second-order rewrite rules \FOLDr
given later in Fig. \ref{fig:FOLDr}.

\subsec{Equational logical framework for cyclic computation}
How to understand the meaning of the result \code{cy(x.S(S(S(x))))} is
arguable.
The overall situation we have demonstrated is illustrated in Fig. \ref{fig:clist-view}.
In this paper, we also provide a formal basis to understand and to
reason about cyclic data, as well as computation results.
We use second-order equational logic
and the axioms \AxCy to equate cyclic data formally
(Fig. \ref{fig:clist-view} [II]).
It completely characterises
the notion of bisimulation on cyclic data.
We can formally prove that an equation on cyclic data, such as
\begin{equation}
\code{cy(x.S(S(S(x))))} \;=\; \code{cy(x.S(x))}
\end{equation}
is derivable from the axioms \AxCy
in the second-order equational logic.
Since \AxCy characterises bisimulation,
it means that the expression \code{cy(x.S(S(S(x))))} is 
bisimilar to \code{cy(x.S(x))}, which is 
a minimal representation of the result regarded as
$\infty$ (infinity).
In this paper, we do not develop an explicit algorithm to extract 
such a minimal result from the computation result, but it is noteworthy that
this equational theory generated by \AxCy is decidable \cite{BE}.
Consequently, it is computationally reasonable.
More practical examples on cyclic datatypes
and computation will be given in \Sec \ref{sec:prog}.

\subsec{Proof by rewriting}
In \Sec \ref{sec:CR},
we will develop a decidable proof method for equations involving
functions defined by fold. 
We give an algorithm to prove whether 
an equation, e.g.
\begin{equation}\label{eq:sampleEq}
  \code{sum(cy(x.2::1::x))}  \;=\;  \code{plus( sum(cy(x.4::5::x)), cy(x.x) )}
\end{equation}
holds or not under \AxCy.
The algorithm is first rewriting both sides of the equation to normal forms,

\begin{diagram}[height=1.7em]
\qquad\code{sum(cy(x.2::1::x))} &\quad\quad& \code{plus( sum(cy(x.4::5::x)), cy(x.x) )} \\
 &&\dTo>* \\
\dTo<*   & & \code{plus(cy(x.S$^9$(x)), cy(x.x))} \\
   & &\dTo>* \\
\code{cy(x.S$^3$(x))} &\bisim& \code{cy(x.cy(x.S$^{9}$(x)))}
\end{diagram}
\medskip

\noindent
then comparing the normal forms by the bisimulation \bisim on cyclic data.
In this case, these are actually bisimilar, hence we conclude
(\ref{eq:sampleEq}) holds.

Why is this methodology correct?
In this paper, we give rigorous reasons and proofs 
validating the methodology in \Sec \ref{sec:CR}.
An important fact is that even when an equation involving cyclic data
(as the above example)
the method is ensured to be decidable.
The rewrite relation ``$\to$'' has two important properties:
strong normalisation (\Sec \ref{sec:SN}) and 
Church-Rosser modulo bisimulation (\Sec \ref{sec:CR}),
both of which are necessary to establish this proof method.
We will prove these rewriting properties by employing advanced rewriting
techniques known as \GS \cite{IDTS00,Blanqui-TCS}, 
and local uniform coherence \cite{JouannaudKRijcai83}.

Note that one can also show that
an equation on cyclic data, such as (\ref{eq:sampleEq}), holds using
ordinary domain theoretic semantics of lazy functional programs.
A remarkable fact is that using our framework,
we can show it without using 
domains, complete partial orders, or complex proof methods.
Our methodology is simpler and decidable, i.e.,
strongly normalising
rewriting and comparison by bisimulation.

\subsection{Previous approaches: cyclic structures %as terms with variable binding in
and functional programming}

Fegaras and Sheard \cite{DBLP:conf/popl/FegarasS96} 
established initial algebra semantics of
mixed variant recursive types, and showed that
cyclic structures (without any quotient) can be encoded by variable binding in
higher-order abstract syntax (HOAS).
It was a starting point of 
succeeding works \cite{TFPcyc,tlca,LMCS,GrICFP,FLOPS} 
improving their HOAS encoding (which had a few drawbacks)
of cyclic structures.

Our previous four papers \cite{TFPcyc,tlca,LMCS,FLOPS} 
improved and extended it to various directions.
Especially, \cite{TFPcyc, tlca, LMCS}
aimed to capture unique representations of cyclic sharing data structures
(without any quotient) in order to obtain an efficient functional programming concept.
In the present paper, 
we will assume the axioms \AxCy and \AxBr (Fig. \ref{fig:axioms})
to equate bisimilar graphs.
Uniqueness of representations is desirable, 
but uniqueness \W{up to bisimilarity} chosen in this paper is 
also desirable,
because checking bisimilarity is efficiently decidable \cite{eff-bisim}
and reflects the meaning of cyclicity.

Oliveira and Cook \cite{GrICFP} also improved \cite{DBLP:conf/popl/FegarasS96} 
by using the parametric HOAS encoding \cite{PhoasICFP08} 
of variable binding for cycle constructs (without any quotient), instead of HOAS,
and developed %various practical 
generic 
fold
combinators on them for functional 
programming.

In all the works mentioned in this subsection, bisimilarity was not
used for identifications of cyclic structures.

\subsection{Our recent work}
In \cite{FICS,MSCS}, the author and collaborators gave algebraic and categorical semantics 
of a graph transformation language UnCAL \cite{Buneman,ICFP10}
using iteration theories \cite{BE}.
The graph data of UnCAL corresponds to cyclic sharing trees of type 
\code{CTree} in the present paper, where graphs are treated 
modulo bisimulation.
UnCAL does not have the notion of types, thus structural recursive 
functions in UnCAL are
always transformations from general graphs to graphs,
thus typing such as \code{sum:CList}\tto\code{CNat} (in Introduction)
or \code{collect:FriendGraph}\tto\code{Names} (in \Sec \ref{sec:prog}) 
could not be formulated.
The present paper develops a suitable algebraic 
framework that captures
datatypes supporting cycles and sharing. 

\subsection{Novelty of this paper}

Since cyclic data structures is potentially dangerous because of cyclicity
as exemplified in the begining of this section,
termination of fold (or iterator) on cyclic structures should be ensured.
However, \W{none of the previous works} including %ours 
\cite{DBLP:conf/popl/FegarasS96,TFPcyc,tlca,LMCS,FLOPS,GrICFP,Buneman,ICFP10,MA}
\W{on cyclic structures 
have formally proved termination nor strong
normalisation of fold},
and the preservation of suitable equivalence on cyclic structures by fold.
Moreover, these works did not developed
a \W{decidable proof method} to prove equations on cyclic data with fold,
such as the equation (\ref{eq:sampleEq}).
In contrast to it, we will prove and provide a decidable proof method by showing
rewriting properties of \FOLDr, i.e.,
strong normalisation and Church-Rosser modulo bisimulation,
which are new results.

\subsec{Organisation} The paper is organised as follows.
We first introduce 
cartesian second-order algebraic theories, which give syntax
and  equational logic of cyclic datatypes in \Sec \ref{sec:salg}.
We next give a categorical semantics of cyclic datatypes in 
\Sec \ref{sec:cat}.
We then extract the fold function from the categorical semantics
in \Sec \ref{sec:fold}.
In \Sec \ref{sec:SN}, we extract second-order rewrite rules of fold and 
show strong normalisation.
In \Sec \ref{sec:CR}, we prove Church-Rosser modulo bisimulation
of the rewrite system of fold and then obtain decidability of the equational 
theory.
In \Sec \ref{sec:prog},
we consider several examples of computing by fold on cyclic datatypes.
In \Sec \ref{sec:conc}, we summarise the paper and discuss related work.

This paper is the fully reworked and extended version of the conference paper
\cite{fscd-cyc}. Besides proofs of all results, the present paper establishes Church-Rosser modulo bisimilarity (Thm. 6.8) and whence decidability of the theory of cyclic data and fold (Cor. 6.10). In addition, there is now a precise description of how primitive recursive functions are defined in our framework (\Sec 4.4), and additional examples are provided (\Sec 7). 

% Local Variables:
% TeX-master: "fs"
% End:

\section{Second-Order Algebraic Theory of Cyclic Datatypes}\label{sec:salg}

We introduce the framework of 
second-order cartesian algebraic theory, which is a typed
and cartesian extension of second-order equational logic in
\cite{2ndCSL} and \cite{HSL}.
Here ``cartesian'' means that the target sort of a function symbol is 
a sequence of base types and we allow unrestricted substitution for variables.
We use second-order algebraic theory as a formal framework
to provide syntax and to describe axioms of algebraic datatypes
enriched with cyclic constructs.
The second-order feature is necessary for the cycle construct and the fold
function on them. We will often omit superscripts or subscripts
of a mathematical object if they are clear from contexts.
We use the vector notation $\vec{A}$ for a sequence $A_1,\ccc,A_n$, 
and $|\vec{A}|$ for its length.

\subsection{Cartesian Second-Order Algebraic Theory}\label{sec:soa}
We assume that
\BB is a set of \Hi{base types} (written as $a,b,c,\ooo$),
and
\Sig, called a \Hi{signature}, is
a set of function symbols of the form
$$f: (\vec{a_1}\to \vec{b_1}),\ooo,(\vec{a_m}\to \vec{b_m}) \to c_1,\ooo,c_n.$$
where all $a_i,b_i,c_i$ are base types
(thus any function symbol is of up to second-order type).
A sequence of types may be empty in the above definition. 
The empty sequence is denoted by $\emptytype$, which may be omitted,
e.g., $b_1,\ooo,b_m \to c\, ,$ or
$\emptytype\to c$. The latter case is simply denoted by $c$.
A signature $\Sig_c$ for type $c$ denotes a subset of \Sig, where every function symbol
is of the form $f: \vec\tau \to c$, which is regarded as a constructor of $c$.
A \Hi{metavariable} is a variable of (at most) first-order type,
declared as
$
\va m : \vec a \to b
$
(written as small-caps letters $\va t, \va s,  \va m, \ooo$).
A \Hi{variable} of a base type 
is merely called variable
(written usually $x,y, \ooo$, or sometimes written
$x^b$ when it is of type $b$).
The raw syntax is given as follows.\medskip
\[
\begin{array}{lllllll}
\text{- \Hi{Terms} have the form}
&
t ::= x \| {x}.{t} \| f(t_1,\ooo,t_n).

\\
\text{- \Hi{Meta-terms} extend  terms to}
&
t ::= x \| {x}.{t} \| f(t_1,\ooo,t_n) \| \m{\va{m}}{t_1,\ooo,t_n}.
\end{array}
\]

\noindent
The last form is called a \W{meta-application}, meaning that 
when we instantiate $\va m : \vec a \to b$ with a term $s$,
free variables of $s$ (which are of types $\vec a$)
are replaced with (meta-)terms $t_1,\ooo,t_n$ (cf. Def. \ref{def:msubst}).
We may write $x_1,\ooo,x_n.\,t\,$ for
$x_1.\ccc.x_n.\,t$, and we assume ordinary \alpha-equivalence 
for bound variables.
Terms are used for representing
concrete cyclic data, functional programs on them and 
equations we want to model. A second-order equational theory 
is a set of proved equations built from terms (N.B. not meta-terms).
Meta-terms are used for formulating equational axioms, which are expected
to be instantiated to terms.

A metavariable context $\Theta$ is a sequence of (metavariable:type)-pairs,
and
a context \Gamma is a sequence of (variable:base type)-pairs.
A judgment is of the form
$$
\mju{\Theta}\Gamma t {\vec b}.
$$
If $\Theta$ is empty, we may simply write $\jud\Gamma t {\vec b}.$
A meta-term $t$ is well-typed by the typing rules
Fig. \ref{fig:meta-terms}.
We omit often the types for binders
as 
$f(\;{\vec{x_1}}.{t_1}, \; \ooo, \; 
    {\vec{x_n}}.{t_n} \;)$.
Given a meta-term $t$ with free variables $x_1,\ooo,x_n$,
the notation $t\subi{x_1\mapsto s_1,\ooo,x_n\mapsto s_n}\; $
denotes ordinary capture avoiding substitution that replaces
the variables with terms $s_1,\ooo,s_n$.

\DefTitled[def:msubst]{Substitution of terms for metavariables}
Let $\Gamma={y_1\:\vec{b_1}\,\ccc,y_k\:\vec{b_1}}$.
Suppose 
\begin{meqa}
\jud {\Gamma',\, x_i^1\:a_i^1,\ooo,x_i^{n_i}\:a_i^{n_i}
}{& s_i}{\vec{b_i}}\qquad (1\le i\le k)
\\
\mju{\va m_1 :{\vec{a_1}\to \vec{b_1}},\ooo,
   \va m_k :{\vec{a_k}\to \vec{b_k}}}{\Gamma} {&e} \vec c %\\
\end{meqa}
where $n_i=|\vec{a_i}|$ and 
$\vec{a_i} = a_i^1,\ooo,a_i^{n_i}$.
Then a substitution
 $\;\jud  {\Gamma,\Gamma'} {e \mesub} \vec c\;$
is inductively defined as follows.%\y{-1em}
\[
\arraycolsep = 1mm
\begin{array}[h]{rclllllllllllllllllll}
x \mesub &\deq& x \\
\va m_i[t_1,\ooo,t_{n_i}] \mesub &\deq& 
  s_i  \;\{{x_i^1} \mapsto t_1\mesub,\ooo,
{x_i^{n_i}} \mapsto t_{n_i}\mesub \} \\
f(\vec{y_1}.t_1,\ooo,\vec{y_m}.t_m) \mesub &\deq& 
f(\vec{y_1}.t_1 \mesub,\ooo,\vec{y_m}.t_m \mesub)
\end{array}
\]
where $\mesub$ denotes $[\va m_1 := s_1,\ooo,\va m_k := s_k]$.
\oDef

\Remark\label{rem:metaterms}
The syntactic structure of meta-terms and substitution
for abstract syntax with variable binding
was introduced  by  Aczel \cite{Aczel}.  This  formal
language  allowed  him  to  consider  a  general  framework  of
rewrite rules for calculi with variable binding. This influenced
Klop's rewrite system of Combinatory  Reduction  System (CRS) \cite{Klop}.
A second-order substitution in the sense of Courcelle \cite{InfTree} 
is very similar to a substitution for metavariables 
but there are no variable binders in the language used in \cite{InfTree}
where function symbols are the targets of replacements
(instead of metavariables in our framework).
\oRemark

\begin{rulefigw}%\y{-.5em}
\begin{meq}
\nxinfrule{Name}
{y : b\in \Gamma}
{\mju \Theta \Gamma y b}
\\[.5em]
\nxinfrule{Env}
{ 
\begin{array}[h]{lllll}
(\va{m}:a_1,\ooo,a_m\to \vec b) \in \Theta \quad\\
 \mju \Theta \Gamma {t_i}{a_i} \quad 
 (1\le i \le m)
\end{array}
}
{\mju \Theta \Gamma {\va{m} [t_1, \ooo, t_m]} \vec{b} }
\qquad
\nxinfrule{Fun}
{
\begin{array}[h]{lllll}
f : (\vec{a_1}\to\vec{b_1}),\ccc,(\vec {a_m}\to \vec{b_m})\to \vec c \in \Sig
\\
\jud {\Gamma,\vec{x_i : a_i}} {t_i} {\vec {b_i}} \quad 
 (1\le i \le m)
\end{array}
}
{ \mju{\Theta}{\Gamma} {f(\; \vec{x_1^{a_1}}.t_1, \ooo \vec{x_m^{a_m}}. t_m \;)
    }{\vec c}
}
\end{meq}
\caption{Typing rules of meta-terms}
\y{.5em} 
\hrule height 0.5pt
\y{.5em} 
\label{fig:meta-terms}

\begin{meq}
\arraycolsep = 0mm
\ninfrule{Ax1}
{
\begin{array}[h]{lllll}
\jud{\Gamma',\vec{x_i : a_i}} {s_i} {\vec {b_i}} \quad 
 (1\le i \le k)
\\
(\mju {\va m_1 :{\vec {a_1}\to \vec {b_1}},\ooo,\va m_k :
  {\vec {a_k}\to \vec{b_k}}}
\Gamma{t_1=t_2} {\vec c} )\in \EE
\end{array}
}
{\jud {\Gamma,\Gamma'} {t_1 \mesub = t_2 \mesub } {\vec c}}
\\[.7em]
\ninfrule{Ax2}
{
\begin{array}[h]{lllll}
\jud {\Gamma',\vec{x_i : a_i}} {s_i} {\vec {b_i}} \quad 
 (1\le i \le k)
\\
(\mju {\va m_1 :{\vec {a_1}\to \vec {b_1}},\ooo,\va m_k :
  {\vec {a_k}\to \vec{b_k}}}
\Gamma{t_1=t_2} {\vec c} )\in \EE
\end{array}
}
{\jud {\Gamma,\Gamma'} {t_2 \mesub = t_1 \mesub } {\vec c}}
\\[.7em]
\ninfrule{Fun}
{
\begin{array}[h]{lllll}
f : (\vec{a_1}\to\vec{b_1}),\ccc,(\vec {a_k}\to \vec{b_k})\to \vec c \in \Sig
\\
\jud {\Gamma,\vec{x_i : a_i}} {t_i=t'_i} {\vec {b_i}} \quad 
 (\text{some }i\text{ s.t. }1\le i \le k)
\end{array}
}
{\jud {\Gamma} 
{f(\vec{x_1^{a_1}}.t_1,\ooo,\vec{x_i^{a_i}}.t_i,\ooo\vec{x_1^{a_1}}.t_k) = f(\vec{x_1^{a_1}}.t_1,\ooo,\vec{x_i^{a_i}}.t'_i,\ooo,\vec{x_1^{a_1}}.t_k) } {\vec c}}
\\[.7em]
\aninfrule{Ref}
{\phantom{\jud\Gamma{t = t}{\vec c}}}
{\jud\Gamma{t = t}{\vec c}}
\qquad
\aninfrule{Tra}
{\begin{array}[h]{lllll}
\jud\Gamma{s = t}{\vec c}\\ %\infspc 
\jud\Gamma{t = u}{\vec c}
\end{array}
}
{\jud\Gamma{s = u}{\vec c}}
\qquad
\aninfrule{OSub}
{
\begin{array}[h]{lllll}
\jud{\Gamma} {s_i} {\vec {b_i}} \quad  (1\le i \le k)
\\
\jud {{x_1 \: b_1,\ooo, x_k\: b_k}} {t=t'} {\vec {c}}
\end{array}
}
{\jud {\Gamma} {t\, [\,\vec{x\mapsto s}\,] = t'\, [\,\vec{x\mapsto s}\,]} {\vec c}}
\end{meq}

\begin{minipage}[t]{\textwidth}\small
In (Ax1)(Ax2), $\mesub$ denotes $[\,\va m_1:= s_1,\ooo,\va m_k:= s_k \,]$.
\\
In (OSub), $[\,\vec{x\mapsto s}\,]$ denotes $[\,x_1\mapsto s_1,\ooo,x_k\mapsto s_k \,]$.
\end{minipage}

\caption{Cartesian second-order equational logic}
\label{fig:sel}
\end{rulefigw}

\bigskip
For meta-terms $\mju{\Theta}{\Gamma}{ s }{\vec b}$ and 
$\mju{\Theta}{\Gamma}{ t }{\vec b},$ an \Hi{equation} is of the form
$$
\mju \Theta \Gamma {s = t} {\vec b},
$$
or denoted by $\jud \Gamma {s = t} {\vec b}$
when \Theta is empty.
The cartesian second-order equational logic is a logic
to deduce formally proved equations from a given set \EE of equations,
regarded as \Hi{axioms}.
The inference system of equational logic is given in Fig. \ref{fig:sel}.
Note that the symmetry rule 
\begin{meq}
\ninfrule{Sym}
{\jud\Gamma{s = t}{\vec c}}
{\jud\Gamma{t = s}{\vec c}}
\end{meq}
is derivable
because of symmetry of (Ax1) and (Ax2).

\subsec{Preliminaries for datatypes}
The \Hi{default signature} \SigGen %for default constructors
is given by the following function symbols called \Hi{default constructors}:
\[
\arraycolsep = .1mm
\begin{array}[h]{llllrlllll}
\RM{Empty sequence}\enspace\ &\EMP &:& \emptytype
&\RM{Tuple}\;\ <-,\ccc,-> &:& (\vec {c_1}),\ooo,(\vec {c_n})\to \vec {c_1},\ooo,\vec {c_n}
\\
\RM{Cycle constructor}&\cyc^{|\vec c|} &:&(\vec c\to \vec c)\to \vec c\enspace \
&\RM{Composition}\;\  \at_{(\vec a,\vec c)} &:& (\vec a\to \vec c),\vec a \to \vec c 
\end{array}
\]

\noindent
defined for all base types
$\vec a, \vec c, \vec {c_1},\ooo, \vec {c_n} \in \BB$. 
This means that any base type has default constructors.
We assume that any signature \Sig includes \SigGen in this paper.
A \Hi{datatype declaration} for a type $c$
is given by a triple $$(c,\Sig_c,\EE)$$ consisting of a base type $c$,
signature $\Sig_c$ and axioms $\EE$, where
every $f\in\Sig_c$ is of the form
$$f : \vec a,c,\ooo,c\to c,$$ 
where $\vec a$ are types other than $c$ (which may be empty),
and any equation in $\EE$ is built from 
$\Sig_c$-terms.

\begin{rulefigw}\yy{-1em} 
\begin{minipage}[t]{\textwidth}
\textbf{Axioms \AxG for cycles}%\y{-.5em}
\end{minipage}
\[\small
\renewcommand{\arraystretch}{1.2}
\arraycolsep = .3mm
\begin{array}[h]{lclrcllllllll}\small
\urule{(sub)}&
\aju{
  \renewcommand{\arraystretch}{.5}
  \begin{array}[h]{llllll}
&\va t \:\!\vec a\to\vec c,\\
&\va s_1,\ooo,\va s_n \: \vec a
  \end{array}
}
{(\vec y.\va t[\vec y])\, \at <\va s_1\opl\ooo\opl \va s_n> &=& \va t[\va s_1\opl\ooo\opl \va s_n]}
\vec c
\\[1em]
\urule{(SP)} &  
\aju{\va t:\uar \vec c}
{\pa{(\vec y.y_1) \at \va t,\ooo, (\vec y.y_n) \at \va t} &=& \va t }
\vec c 
\\[1em]
\urule{(dinat$_1$)} &
  \renewcommand{\arraystretch}{.5}
  \aju{
  \begin{array}[h]{lll}
    \va s\: a\!\to\! c,\\ \va t\: c\!\to\! a
  \end{array}
  }
{\cy{ x. \va s[\va t[ x]]} &=& \va s[ \cy{  z.\va t [\va s[ z]] }] ] }
 c 
\\[1em]
\urule{(dinat$_n$)} &
  \renewcommand{\arraystretch}{.5}
  \aju{
  \begin{array}[h]{lll}
    \va s\:\vec a\!\to\!\vec c,\\ \va t\:\vec c\!\to\!\vec a
  \end{array}
  }
{\cy{\vec x. \va s[\va t[\vec x]]} &=& \va (\vec z.\va s[\vec z]) \at  \cy{ \vec z.\va t [\va s[\vec z]] } ] }
\vec c 
\\[1em]
\urule{{(\Bekic)}} & 
\aju{
  \renewcommand{\arraystretch}{.5}
  \begin{array}[h]{lll}
  \va t: \vec c,\vec a\to\vec c,\\
  \va s:\vec c,\vec a\to\vec a
  \end{array}
 }
{
\multicolumn{3}{l}{
\cy{\vec x,\!\vec y\!.\,<\,\btt,\; \bss\,>}
=\!
\begin{array}[h]{lll}
<\;\  &\cy{\vec x.\, (\vec y.\btt) \at    \cy{\vec y. \bss}     },\\
  &\cy{\vec y.\, (\vec x.\bss) \at  \cy{\vec x.\, (\vec y.\btt) \at  \cy{\vec y. \bss}}
     }\;>
\end{array}
}
}
{\vec c,\vec a}
\\[1em]
\urule{(CI)} & 
\aju{\va t:a^m \to a}
{\cy{\vec y. \pa{ \va t[\rho_1],\ooo, \va t[\rho_m] }}
&=& <\cy{y.\bbtt},\ooo,\cy{y.\bbtt}>}
a^m
\\[.5em]
 %{\text{These are defined for any base types }\vec b,\vec a,\vec c.}\\
\end{array}
\]

\begin{minipage}[t]{\textwidth}\small
In (dinat$_n$), $|\vec c|=m\gt 1$ and
$\va s[\va t[\vec x]]$ is short for 
$\va s[(\vec y.y_1) \at \va t[\vec x],\ooo,
   (\vec y.y_m) \at \va t[\vec x]]$.
Similarly for
$\va t[\va s[\vec z]]$.
In (\Bekic), 
$\btt$ and $\bss$ are short for $\va t[\vec x,\vec y]$ and $\va s[\vec x,\vec y]$,
respectively.
In (CI), 
$
\rho_i= y_{i_1},\ooo\opl y_{i_m},
$
where $i_1,\ooo,i_m \in \set{1,\ooo,m}$,
$\bbtt$ is short for 
$\va t[y,\ooo,y]$.
\end{minipage}

\bigskip
\hrule height 0.5pt

\bigskip
\begin{minipage}[t]{\textwidth}\normalsize
\textbf{Axioms $\AxBr(\vnil,+)$ for branching}%\y{-.5em}
\end{minipage}
\[\normalsize
\renewcommand{\arraystretch}{1}
\begin{array}[h]{lclrcllllllll}\normalsize
\urule{(del)} & 
\aju{\va t: c}
{\cy{x^c.\Uni{x}{\va t}} &=& \va t}
c \\
\urule{(unitL)}& \aju{\va t: c}
{\Uni{\vnil}{\va t} &=& \va t }
c \\
\urule{(unitR)}& \aju{\va t: c}
{\Uni{\va t}{\vnil} &=& \va t }  
c \\
\urule{(assoc)}& \aju{\va s,\va t,\va u: c}
{\Uni{(\Uni{\va s}{\va t})}{\va u} &=& \Uni{\va s}{(\Uni{\va t}{\va u})}} 
c \\
\urule{(comm)}& \aju{\va s,\va t: c}
{\Uni{\va s}{\va t} &=&  \Uni{\va t}{\va s} }
c \\
\urule{(degen)}& \aju{\va t: c}
{\Uni{\va t}{\va t} &=&  \va t }
c 
\end{array}
\]
\medskip

\begin{minipage}[t]{\textwidth}\small
Note that the axiom $\AxBr(\vnil,+)$ is parameterised by
the function symbols $\vnil,+$.
In general, writing 
$\AxBr(\nu,\mu)$ where $\nu : c,\; \mu:c,c\to c \in\SigCon$,
we mean the set of axioms obtained from the above axioms by replacing
$\vnil$ with \nu, and $+$ with \mu.
\end{minipage}
\caption{Axioms} 
\label{fig:axioms}
\end{rulefigw}

\subsection{Instance (1): Cyclic Lists modulo Bisimulation}
\label{sec:cytype}

We will present an algebraic formulation of cyclic datatypes.
By a cyclic datatype, we mean an algebraic datatype having 
the cycle construct \fcy
satisfying the axioms that characterise cyclicity.
The first example is the datatype of natural numbers.
It has already been defined as \code{CNat} in Introduction as a pseudo code.
We now give a formal definition using a datatype declaration.
The datatype declaration for \CNat is given by
$(\CNat, \Sig_\CNat, \AxCy)$ where $\Sig_\CNat$ is 
$$%\y{-.5em}
\code 0 :\CNat, \qquad 
\code S : \CNat \to \CNat
$$
and the axioms \AxCy are given in Fig. \ref{fig:axioms}.

The second example is the datatype of cyclic lists.
It has already been defined as \code{CList} in Introduction as a pseudo code.
Fix $a\in\BB$.
The datatype declaration for \Lista, the cyclic lists of type $a$, is given by
$(\Lista, \Sig_\Lista, \AxCy)$ where $\Sig_\Lista$ is given by
$$%\y{-.5em}
\vnil :\Lista, \qquad 
({-}\coSym_a{-}) : a, \Lista \to \Lista
$$
and the axioms \AxCy are given in Fig. \ref{fig:axioms}.
Note that \Lista has also the default constructors,
thus one can form a cycle (see the example below).
The definition of \code{CList} in Introduction actually represents
the datatype declaration 
$(\code{CList}_\code{CNat}, \Sig_{\List_\code{CNat}}, \AxCy)$.
Hereafter, we often omit the type parameter subscript $a$ of \code{CList}.
The axioms \AxG  mathematically characterise
that \fcy is truly a cycle constructor in the sense of 
Conway fixed point operator \cite{BE}. The equational theory generated by 
\AxG captures the intended meaning of cyclic lists.
For example, the following are identified as the same cyclic list:

\bigskip
\begin{center}
\ \qquad\qquad\includegraphics[scale=.45]{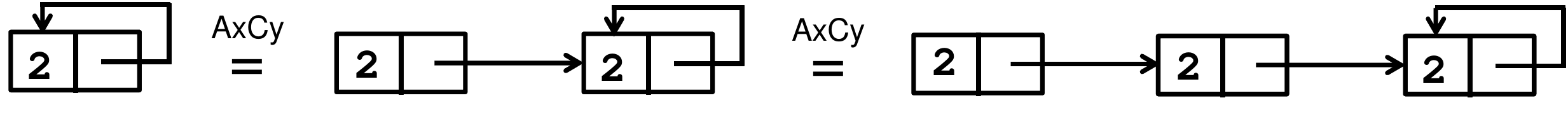}  \\
\end{center}
$$
\cy{x. 2 \co x} \quad=\quad  2 \co \cy{x. 2 \co x} \quad=\quad 2 \co 2 \co \cy{x. 2 \co x}
$$
These equalities come from the \W{fixed point law} of \fcy:
  \[
\begin{array}[h]{lrcllllllll}
\urule{(fix)} &
  \renewcommand{\arraystretch}{.5}
\cy{ x. \va s[ x]} &=& 
\va s[\; \cy{ x.\va s[ x]] } \;]
\end{array}
\]
which is an instance of the axiom (dinat) when $\va t$ is identity
(i.e., $\va t \mapsto x$ for $\jud{{x\: c}}{x} c$).

\subsec{On axioms \AxCy}
We explain the intuitive meaning of the axioms in \AxCy. %in further detail.
Parameterised fixed-point axioms axiomatise 
\fcy as a fixed point operator.
They (minus (CI)) are equivalent to the axioms for
Conway operators of %Bloom and \Esik 
\cite{BE,Hassei,alex-plot}.
Beki\u{c} law is well-known in denotational semantics 
(cf. \cite[\Sec 10.1]{Winskel}) to calculate the fixed point of a pair
of continuous functionns.
Conway operators are also arisen in work independently 
of Hyland and Hasegawa
\cite{Hassei}, who established a connection
with the notion of traced cartesian categories \cite{JSV}.
There are equalities that holds in the cpo semantics but Conway operators
do not satisfy.
The axiom (CI) is the commutative identities of Bloom and \Esik 
\cite{BE,alex-plot},
which ensures that %such 
all equalities that hold in the cpo semantics
do hold.
See also \cite[Section 2]{alex-plot}  for a useful overview about
this.
The equality generated by \AxCy is bisimulation on cyclic lists.
This is included in the equality of cyclic sharing trees
given in the next subsection.

\subsection{Instance (2): Cyclic Sharing Trees modulo Bisimulation}
\label{sec:ctree}
Next we consider the datatype of 
binary branching trees, which can involve cycle and sharing.
We call them \W{cyclic sharing trees}, or simply cyclic trees.
We first give the declaration of datatype \code{CTree} of cyclic trees 
as the style of pseudo code below, where
we assume that 
$f_1,\ooo,f_n$'s part denotes various unary function symbols such as \code{a,b,c,p,q,}$\ooo$.

\begin{CVerbatim}[commandchars=\\\{\},codes=\mathcom]
ctype CTree where
  $\fone$ : CTree \tto CTree     
   $\vdots$  
  $\fenu$ : CTree \tto CTree     
  $\vnil\;$ : CTree
  +\;\, : CTree,CTree \tto CTree
with axioms \nAxCy,\nAxBr(\nil,+) 
\end{CVerbatim}
Formally, it is expressed as the datatype declaration
$$(\iota,\, \set{f_1,\ooo,f_n, \vnil, + },\;\AxCy\union\AxBr(\vnil,+)).$$
The binary operator \code{+} denotes a branch.
For example, one can write \code{b([])+c([])} (cf. Fig. \ref{fig:ex} (A)).
The datatype can express sharing by the constructor $\at$
of composition:
\begin{CVerbatim}[commandchars=\\\{\},codes=\mathcom]
(x.a(b(x) + c(x))) $\at$ p([])
\end{CVerbatim}
(cf. Fig. \ref{fig:ex} (F)).
Note that the first argument of composition $\at$ has 
a binder (e.g. \code{x.}),
which indicates a placeholder filled by the shared part after $\at$ (e.g. \code{p([])}).
A binder at the first argument of $\at$-term may be a sequence of variables 
(e.g. ``\code{x,y.}'' in (E)), which will be filled by terms in a tuple 
(e.g. \code{<p([]),q([])>}).
Cyclic trees are very expressive.
They cover essentially XML trees with IDREF, 
the data model called \W{trees with pointers} \cite{contextlogic},
and arbitrary rooted directed graphs (cf. Fig. \ref{fig:ex} (B)(E)).

We denote by \bisim the equivalence relation
generated by the axioms $\nAxCy,\AxBr(\vnil,+)$ in Fig. \ref{fig:axioms}.
Using the axioms $\AxCy\union\AxBr(\vnil,+)$,
we can reason this equality \bisim in the second-order equational logic.
The equality \bisim gives reasonable meaning of cycles as in the case of cyclic lists.
The branch $+$ is associative, commutative
and idempotent (cf. Fig. \ref{fig:ex} (C)), thus nested $+$ can be
seen as an $n$-ary branch (cf. Fig. \ref{fig:ex} (D)).
Moreover, a shared term and its unfolding are also identified by \bisim
because of the axiom (sub) (cf. Fig. \ref{fig:ex} (F)).
The axiom (sub)
is similar to the \beta-reduction in the \lmd-calculus.

\begin{rulefigt}
\includegraphics[scale=.51]{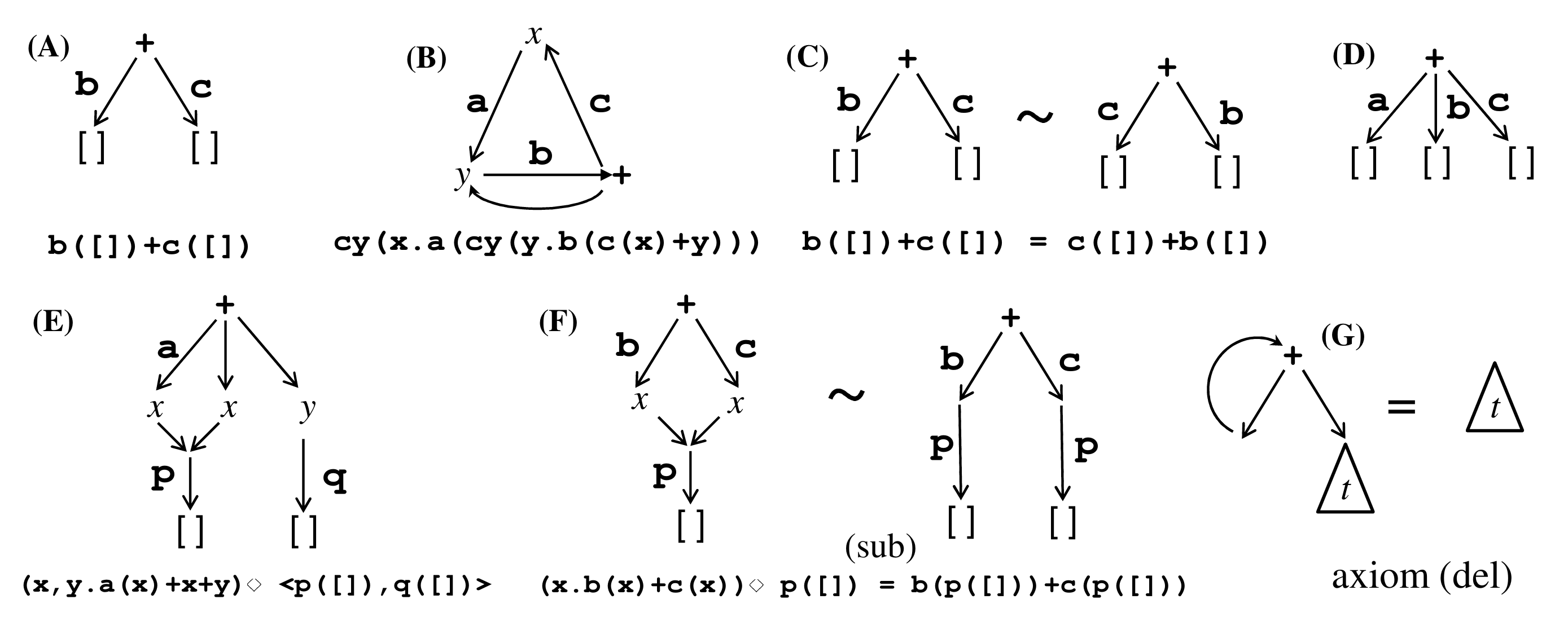}
\caption{Examples of cyclic sharing trees}
\label{fig:ex}
\end{rulefigt}

\subsec{Algebraic theory of bisimulation}
Actually, \bisim is exactly \Hi{bisimulation} on cyclic trees.
Since unary $f$ expresses a labelled edge, and $+$ expresses a branch,
{cyclic sharing trees} are essentially 
process graphs of regular behaviors,
called \W{charts} by Milner in \cite{MilnerRegular}.
Infinite unfolding of them are synchronization trees \cite{BE}.
Thus the standard notion of bisimulation 
between graphs can be defined.
Intuitively, starting from the root, bisimulation is by
comparing traces of labels of two graphs along edges 
(more detailed definition is given in \cite{BE,MilnerRegular} or 
(\cite{MSCS} Appendix)).
Now we see that (C),(F) and (G) are examples of bisimulation.
Actually, the axioms in Fig. \ref{fig:axioms} are sound for bisimulation, i.e., 
for each axiom, the left and the right-hand sides
are bisimilar. Moreover, it is complete.

\PropTitled[th:bisim-ax]{\cite{FICS},(\cite{MSCS}\Sec 5.3)}
$\jud\Gamma { s = t} {\iota}$ is derivable from \AxG and 
$\AxBr(\vnil,+)$
iff if $s$ and $t$ are bisimilar.
\oProp

The main reason of this is that the axioms \AxG and 
$\AxBr(\vnil,+)$ are a second-order representation of 
Bloom and \Esik's complete equational axioms of bisimulation \cite{BE}.
A crucial fact is that bisimulation $s \sim t$ is decidable 
\cite{BE,Buneman}.
There is also an efficient algorithm for checking bisimulation,
e.g. \cite{eff-bisim}.
Hence, cyclic datatypes with the axioms \AxCy, or 
the axioms $\AxCy\union\AxBr$ are
computationally feasible.
For example, equality 
on cyclic structures such as the one we have seen in Fig. \ref{fig:clist-view}
can be checked efficiently.

There are many other instances of cyclic datatypes, some of which will be 
given in \Sec \ref{sec:prog}.

% Local Variables:
% TeX-master: "fs"
% End:

\section{Categorical Semantics of Cyclic Datatypes}\label{sec:cat}

In this section, we give a categorical semantics of cyclic datatypes.
A reason to consider categorical semantics is
to systematically obtain a ``structure preserving map''
on cyclic datatypes.
We will formulate the ``fold'' function for a cyclic datatype
as a functor on the category for cyclic datatypes 
(Thm. \ref{th:universality} and \Sec \ref{sec:fold}).

Since a cyclic datatype has cycles,
the target categorical structure should have a notion of 
fixed point.
It has been studied in iteration theories of Bloom and \Esik 
\cite{BE}. In particular iteration categories \cite{IterCat}
are suitable for our purpose, which
are \Hi{traced cartesian categories} \cite{JSV,Hassei}
satisfying the \W{commutative identities} axiom \cite{BE}.
We write $\terminal$ for the terminal object, $\X$ for the cartesian product,
$<-,->$ for pairing, and $\dup=<\id,\id>$ for diagonal in a cartesian category.

\DefTitledN[def:iter-cat]{\cite{IterCat,BE}}
A \W{Conway operator} in a cartesian category $\CC$ 
is a family of functions
$
\dagg{-}:\CC(A\X X,X) \to \CC(A,X)
$
satisfying:
\begin{meq}
\dagg{f \circ (g \X \id_X)} = \da f \circ g\;,\qquad
\dagg{\da f} = \dagg{f \circ (\id_A \X \dup)}\;,\\ 
f \circ <\id_A,  \dagg{g\circ <\pi_1, f>} > = \dagg{f \circ <\pi_1, g>}.
\end{meq}
An \Hi{iteration category} is a cartesian category having a Conway operator
satisfying the ``commutative identities'' law \cite{BE}
\[
\da{ \pa{f \circ (\id_A \X \rho_1),\ooo,f \circ (\id_A \X \rho_m)} }
=
\Delta_m \circ \dagg{f \circ (\id_A \X \Delta_m)} : A \to X
\]
where
\begin{itemize}[nosep]
\item ${f}:{A \X X^m}\to{X}$
\item $\Delta_m\deq <\id_X, \ccc, \id_X>:{X}\to {X^m}$ is the diagonal
\item ${\rho_i}:{X^m}\to{X^m}$
such that $\rho_i=<q_{i1},\ooo, q_{im}>$ %is a tuple of projections
where each $q_{ij}$ is one of projections $\pi_1,\ooo,\pi_m:{X^m}\to{X}$ 
(see also \cite{alex-plot}).
\end{itemize}
An \W{iteration functor} between iteration categories is a cartesian functor
that preserves Conway operators.
\oDefTitled

\Remark
It is well-known that there are several equivalent axiomatisations of the
Coway operator, cf. \cite[\Sec 6.8]{BE}\cite[\Sec 7.1]{Hassei} and
\cite{alex-plot}. A frequently used axiomatisation is 
a natural and dinatural operator satisfying
the \Bekic law, which is what the axioms \AxCy in Fig. \ref{fig:axioms} state.
\oRemark

A typical example of iteration category is the category of 
\CPO complete partial orders (cpos) with bottom and
continuous functions \cite{BE,Hassei}, where the least fixed point operator is a
Conway operator.

\Def[def:cat-structure]
Let \Sigma be a signature.
A \Hi{\Sig-structure} $M$ in an iteration category \CC is specified
by giving for each base type $b\in\BB$, an object $\den{b}^M$ 
(or simply written $\den{b}$) in \CC,
and for each function symbol
$f: (\vec{a_1}\to \vec{b_1}),\ooo,(\vec{a_m}\to\vec{b_m}) \to \vec c$,
a function
\begin{equation}
\den{f}^M_A : \CC(A\X\den{\vec{a_1}},\den{\vec{b_1}})\X\ccc\X\CC(A\X\den{\vec{a_n}},\den{\vec{b_n}})
       \rTo\CC(A,\den{\vec{c}})
\end{equation}
which is natural in $A$, where 
$\den{b_1,\ooo,b_n} \deq \den{b_1}\X\ooo\X\den{b_n}$.
Also given a context $\Gamma=x_1:b_1,\ooo,x_n:b_n$,
we set $\den{\Gamma} \deq \den{b_1,\ooo,b_n}$.
The superscript of $\den{-}$ may be omitted hereafter.
\oDef

\subsec{Interpretation}
Let $M$ be a \Sig-structure
in an iteration category \CC.
We give the categorical meaning of a term judgment
${\jud \Gamma t {\vec c}}$ (where there are no metavariables)
as
a morphism $\den{t}^M: \den\Gamma \to \den{\vec c}$ in \CC defined by
\begin{meqa}
\den{\jud\Gamma{y_i}{c}}^M &= \pi_i:\den\Gamma \to \den c\\
\den{\jud\Gamma {f(\; \vec{x_1^{a_1}}.t_1, \ooo \vec{x_n^{a_n}}. t_n
    \;)}\vec c}^M
&= \den{f}^M_{\den\Gamma}(\, %\den{\jud{\Gamma,\vec{x_1:a_1}}{t_1}{\vec b} }^M\!\!\!,\ooo,\den{t_n}^M\,
\den{\jud{\Gamma,\vec{x_1:{a_1}}}{t_1}{\vec {b_1}}}^M,\ooo,
\den{\jud{\Gamma,\vec{x_n:{a_n}}}{t_n}{\vec {b_n}}}^M
).
\end{meqa}
We assume the following interpretations in any \SigGen-structure:
\[
\arraycolsep = 1mm
\begin{array}[h]{lllllllllllllllllll}
\den{\EMP}^M_A : \CC(A,\terminal)\to\CC(A,\terminal)\\
\den{\EMP}^M_A(t) = t  \\
\den{<-,\ooo,->}^M_A : 
  \CC(A,\den{\vec {c_1}})\X\ccc\X\CC(A,\den{\vec {c_n}})
  \to \CC(A,\den{\vec {c_1}}\X\ccc\X\den{\vec {c_n}})
\\
\den{<-,\ooo,->}^M_A(t_1,\ooo,t_n) = <t_1,\ooo,t_n> \\
\den{\at}^M_A : \CC(A\X \den{\vec a},\den{\vec c})\X\CC(A,\den{\vec a})\to\CC(A,\den{\vec c})\\
\den{\at}^M_A(t,s) = t \circ <\id_A,s> \\
\den{\fcyc}^M_A : \CC(A\X \den{\vec c},\den{\vec c})\to\CC(A,\den{\vec c})\\
\den{\fcyc}^M_A(t) = \da{t}
\end{array}
\]
We say a \Sig-structure $M$ \Hi{satisfies}
an equation $\jud \Gamma {s=t} c$,\; if $\den{s}^M=\den{t}^M$ holds.
Importantly,
every \SigGen-structure satisfies the axioms \AxG because \CC is an iteration category.

\Def
A \Hi{$(\Sig,\EE)$-structure} is a \Sig-structure $M$ in \CC satisfying 
all equations in a set $\EE$ of axioms.
Let $N$ be a $(\Sig,\EE)$-structure in an iteration category \calD.
We say that an iteration functor 
$F:\CC\to\calD$ \Hi{preserves $(\Sig,\EE)$-structures} if 
$F(\den{-}^M)=\den{-}^N$.

A \Hi{$c$-structure} $(M,\alpha)$ 
for a datatype declaration $(c,\Sig_c,\EE)$
consists of a $(\Sig_c,\EE)$-structure $M$ with
a family of morphisms of \CC:
\[
\alpha \;\deq\; (\;\fden{f}^M : \den{b_1}\X\ooo\X\den{b_n}\to \den{c}\;)_{f : b_1,\ooo,b_n\to c \in \Sig_c}.
\]
Note that it induces
the interpretation of every function symbol $f\in \Sig_c$ 
given by
\[ \den{f}_A^M(t_1,\ooo,t_n) = \fden{f}^M \circ <t_1,\ooo,t_n>\] 
for any $A$ in $\CC$.
We say that an iteration functor 
$F:\CC\to\calD$ \Hi{preserves $c$-structures} if 
$F(\den{c}^M)=\den{c}^N$, and 
$F(\fden{f}^M) = \fden{f}^N$ for every $f\in\Sig_c$.
\oDef

\ExampleTitled[ex:lists-cpo]{The datatype $\List_a$ of cyclic lists of type $a$}
We give a $\List_a$-structure 
in the iteration category \CPO via 
the standard initial algebra characterisation of datatypes
in the category \CPObot of cpos and strict continuous functions \cite{Abramsky-Jung}.
We write that 
$1$ is the cpo $\set{\bot}$ (which is the initial object of \CPObot),
where $\bot$ is the least element,
$\oplus$ is the coalesced sum, and $(-)_\bot$ is lifting.

Let $A$ be the interpretation $\den{a}^L$ of a base type $a$ in \CPO.
We construct 
an initial algebra $L$ of the functor $F$ on \CPObot defined by
$
F(X)=1_\bot \oplus(A\X X)_\bot
$
by using the initial algebra construction \cite{Smyth-Plotkin},
i.e., taking the colimit of \omega-chain
\[
1 \rTo^! F(1) \rTo^{F(!)} F^2(1) \rTo^{F^2(!)} \ccc
\]
we have a cpo $L\iso F(L)$, consisting of
finite and infinite possibly partial lists  
with
continuous functions
$
\fn{nil} : 1 \to L$ and
$\fn{cons} : A\X L \to L$. 
The order of the cpo is the usual point-wise ordering
$\bot \sLE t$, and $(t_1,t_2) \sLE (t'_1,t'_2)$ if 
$t_1 \sLE t'_1$ and  $t_2 \sLE t'_2$, etc.
Then we have
a ${\List_a}$-structure $(L,(\fden{\vnil}^L, \fden{\coSym}^L)) $ 
defined by
\[
\arraycolsep = 1mm
\begin{array}[h]{lllllllllllllllllll}
\fden{\vnil}^L &=& \fn{nil}, \qquad
\fden{\coSym}^L &=& \fn{cons}
\end{array}
\]
and all the axioms in \AxCy hold in \CPO, since these are now well-known
propeties of cpos \cite{Winskel,BE,alex-plot}.
Note that this construction is the same as modelling 
the lazy list datatype,
hence this also shows that our framework covers modelling lazy datatypes
as in Haskell.
\oExample

\ExampleTitled{The cyclic tree type \iota}
A \iota-structure in an iteration category \CC is given by a $(\SigGen,\AxCy\union \nAxBr(\vnil,+))$-structure $M$
where $\den{\iota}^M= N$ and
$N$ is a commutative monoid object $(N,\, \unit \!:\! \terminal\to N,\;
\mu \!:\! N\X N \to N)$ in \CC
satisfying 
\[
\arraycolsep = 1mm
\begin{array}[h]{lllllllllllllllllll}
\fden{\vnil}^M &=& \unit, \qquad
\fden{\uniSym}^M &=& \mu.
\end{array}
\]
It satisfies all axioms of $\AxBr(\vnil,+)$.
Note that every \iota-structure is
a \W{degenerated commutative bialgebra} (cf.\ \cite{FiorePROP})
in a cartesian category \CC,
i.e., $N$ forms also a comonoid $(N,!,\Delta)$
that satisfies the compatibility
\[
\Delta \circ \unit = \unit \X \unit,\quad
\dup\o\mu \;=\; (\mu\X\mu) \circ (\id\X {<\pi_2,\pi_1>}\X\id) \at(\dup\X\dup), \quad
\mu\o\dup \;=\; \id.
\]
The last equation is by
$\mu \circ \Delta = \mu \circ <\id, \id> = \mu \circ <\id, \dagg{\mu}>
\stackrel{\text{(dinat)}}=
\dagg{\mu} = \id$.
Thus, a \iota-structure models branch (by \mu)
and sharing by ($\Delta$)
of cyclic sharing trees.
\oExample

We next give a syntactic category and a \Sig-structure to prove 
categorical completeness.
Let $\Sig$ be a signature,
and $\EE$ a set of axioms which is the union of \AxG and 
axioms for all datatype declarations of base types $c$.
Given axioms \EE, all proved equations $\jud\Gamma{s=t}\bb$
(which must be the empty metavariable context)
by the second-order equational logic (Fig. \ref{fig:sel}),
defines an equivalence relation $=_\EE$ on well-typed terms,
where we also identify renamed terms by bijective renaming 
of free and bound variables.
We 
write an equivalence class of terms by $=_\EE$
as
$
\eclas{\jud \Gamma t \vec c}.
$
We define the category \TE of 
term judgments modulo $=_\EE$ by taking
\begin{itemize}[noitemsep]
\item objects: sequences of base types $\vec c$
\item morphisms: $\eclas{\jud \Gamma t \vec c} : \den{\Gamma} \to
  \den{\vec c}$,\;
      the identity: $\eclas{\jud {\vec {x:c}} {<\vec x>} \vec c}$ % : \den\bb \to \den\bb$.
\item composition: 
$\eclas{\jud {\vec{x:b}} s \vec c} \circ \eclas{\jud \Gamma t \vec b} \deq 
\eclas{\jud \Gamma {(\vec x.s)\at t} \vec c}$
\end{itemize}

\Prop[th:cl-cat]
\TE is
an iteration category, %and each $\den{c}^\GE$
and has a $(\Sig,\EE)$-structure \GE.
\oProp
\proof
We define a \Sig-structure \GE by
$\den{c}^\GE \deq c$ for each $c\in\BB$, and 
 %: (\vec{a_1}\to \vec{b_1}),\ooo,(\vec{a_m}\to\vec{b_m}) \to \vec c, 
\begin{meqa}
&\den{f}^\GE_{\vec a} : \TE( (\vec a,{\vec{a_1}}),{\vec{b_1}})
   \X\ccc\X\TE((\vec a,{\vec{a_n}}),{\vec{b_n}})
       \rTo\TE(\vec a,{\vec{c}})\\
&\den{f}^\GE_{\vec a}(\eclas{t_1},\ooo,\eclas{t_n}) \deq 
\eclas{f(\vec{x_1}.t_1,\ooo,\vec{x_n}.t_n)}
\end{meqa}
for $f: (\vec{a_1}\to \vec{b_1}),\ooo,(\vec{a_m}\to \vec{b_m}) \to
\vec c\in \Sig$
and base types $\vec a$.
This is well-defined because $=_\EE$ is a congruence.
We take
\begin{itemize}[noitemsep]
\item terminal object: $\emptytype$
 \qquad$\bullet $\; pair: $<\eclas{s},\eclas{t}> \deq 
\eclas{\ju{\pa{s\opl t}}{\Gamma}{\vec{c_1},\vec{c_2}  }}$
\item product: concatenation of sequences
\item Conway op.: $\dagg{\eclas{\ju t {\Gamma,\vec{x : c}} {\vec c}}}= 
\eclas{\jud {\Gamma} {\cy{\vec{x^c}.t}} {\vec c}}$
\item projections: $\eclas{\jud{x_1:c_1,x_2:c_2}{x_i}{c_i}}$
\end{itemize}

\noindent
Then these data turn $\TE$ into an iteration category,
and moreover, \GE forms a $(\Sig,\EE)$-structure because of the axioms $\EE$
for each $c\in\BB$.
Moreover, \[
(\,\den{c}^\GE\!,\; (f : \den{\vec{b_1}}\X\ccc\X\den{\vec{b_m}}
\to\den{c})_{f: \vec{b_1},\ooo, \vec{b_m} \to c\in \Sig_c}\,)
\] 
is a $c$-structure. Note that $\den{\vec{b_i}} = \vec{b_i}$ in \TE.
\QED

Then $\den{t}^\GE=\eclas{t}$ holds for all well-typed terms $t$.
Using it, we have the following.

\ThTitled[th:cat-comp]{Categorical soundness and completeness}
$\jud \Gamma {s=t}\bb$ is derivable  iff\;\;
$\den{s}_\CC^M = \den{t}_\CC^M$ holds 
for all iteration categories \CC and all $(\Sig,\EE)$-structures in \CC.
\oTh

\smallskip

\Th[th:universality]%
For a $(\Sig,\EE)$-structure $M$ in an iteration category \CC,
there exists 
a unique iteration functor $\Psi^M : \TE \rTo \CC$
that preserves $(\Sig,\EE)$-structures.
Pictorially, 
it is expressed as
the following diagram, where $\Term$ denotes the set of all terms 
(without quotient).
\begin{diagram}[2em]
&  \Term &\rTo^{\den{-}^\GE}& \TE \\
\quad\qquad&\dTo<{\den{-}^M}&\ldTo>{\Psi^M}\\
&\CC
\end{diagram}
\oTh
\proof
We write simply \Psi for $\Psi^M$.
Since $\Psi$ preserves $(\Sig,\EE)$-structures, 
$\Psi(\den{-}^\GE) = \den{-}^M$ holds. Hence
$\Psi(\den{t}^\GE) = \Psi(\eclas{t}) = \den{t}^M $ for any $t$,
meaning that the mapping $\Psi$ is required to satisfy 
\begin{equation}
\arraycolsep = .5mm
\renewcommand{\arraystretch}{1}
\begin{array}[h]{llllll}
  \Psi(& \eclas{\jud \Gamma { y_i} c} &) &=&\; \pi_i \\
  \Psi(& \eclas{\jud \Gamma \EMP ()} &) &=&   !  \\
\Psi(& \eclas{ \jud\Gamma{\pa{ s \opl t }} {\vec{c_1},\vec{c_2}} } &) &=& 
  <\Psi\eclas{\jud\Gamma s \vec{c_1}} , \Psi\eclas{\jud\Gamma t \vec{c_2}}> 
\\
\Psi(& \eclas{ \jud\Gamma{\cy{\vec {x^c}.t}}{\vec c}} &) &=& 
\dagg{\Psi  \eclas{\jud{\Gamma,\vec{x:c}}t {\vec c}}}
\\
\Psi(& \multicolumn{4}{l}{
\eclas{ \jud \Gamma {f(\; \vec{x_1^{a_1}}.t_1, \ooo,
  \vec{x_m^{a_m}}. t_m )}c })} 
\\
&\multicolumn{4}{l}{= 
\den{f}_{\den \Gamma}^M( \Psi \eclas{\jud {\Gamma, \vec{x_1:a_1}} 
{t_1} b_1}, \ooo%)
, 
             \Psi \eclas{\jud {\Gamma, \vec{x_m:a_m}} {t_m} b_m})  
}
\\
\Psi(& \eclas{ \jud\Gamma{(\vec{x^b}.t) \at s} c } &) &=&   
   \Psi\eclas{\jud{\Gamma,\vec{x:b},} t c} \circ <\id_{\den
     \Gamma},\Psi\eclas{\jud\Gamma s {\vec b}}  >
\end{array}
\label{eq:fold-law}
\end{equation}
The above equations mean that $\Psi$ is an iteration functor %a traced cartesian functor
that sends the $(\Sig,\EE)$-structure \GE to $M$. 
Such $\Psi$ is uniquely determined by these equations because 
$\GE$ is a $(\Sig,\EE)$-structure.
\QED

% Local Variables:
% TeX-master: "fs"
% End:

\section{Fold on Cyclic Datatypes}\label{sec:fold}

Fix a cyclic datatype $c$ (say, the type $\List$ of cyclic lists).
By the previous theorem, for a $c$-structure $M$,
the interpretation $\den{-}^M$ determines
a $c$-structure preserving iteration functor $\Psi^M$.
If we take the target category $\CC$ as also \TE,
$M$ should be another cyclic datatype $b$ (say, the \code{CNat} of 
cyclic natural numbers), where
the constructors of $c$ are interpreted as terms of type $b$.
For example, the sum of a cyclic list in Introduction is understood
in this way.
Thus the functor $\Psi^M$ determined by $\den{-}^M$
can be understood as 
a transformation of cyclic data from terms of type $c$ to 
terms of type $b$.

Along this idea, we formulate the fold operation from the 
cyclic datatype $c$ to another cyclic datatype $b$ by the functor $\Psi^M$.
Let $(c,\Sig_c,\EE_c)$ and
$(b,\Sig_b,\EE_b)$ be datatype declarations.
Let $\EE$ be the set of axioms collecting the axioms of all datatype
declarations for types in \BB, which 
includes $\AxG$ (and $\AxBr$ if the datatype suppose it) for 
every datatype.
Hence $\EE \supseteq \EE_b \union \EE_c$.
We define a $(\Sig_c,\EE)$-structure $(M,\alpha)$ in \TE by
\[
  \den{c}^M  =b 
\]
and $\den{a}^M = a $ for $a\not = c$.
We write the arrow part function $\Psi^M$ on hom-sets as the fold, i.e.,
\begin{equation}
\mfold^c_b(\alpha) : \TE(\vec{c},\; c)\rTo\TE(\den{\vec{c}},\; b).
\label{eq:fold}
\end{equation}
where $\vec c = c_1,\ooo,c_n$ (N.B. some of $c_i$ may be $c$).

\subsection{Axiomatising \texorpdfstring{\fold}{fold} as a second-order algebraic theory}

The \mfold is a function on equivalence classes of term judgments 
modulo $\EE$ 
characterised by (\ref{eq:fold-law}). Equivalently, we regard it
as a function on terms (or term judgments) that preserves $=_\EE$, i.e.,
$$
s =_\EE t \quad\TO\quad \mfold^c_b(\alpha)(s)\; =_\EE\; \mfold^c_b(\alpha)(t).
$$
In this subsection, we axiomatise 
the function $\mfold^c_b$ as the laws of fold
within second-order equational logic using (\ref{eq:fold-law}).

\subsubsection*{Axiomatising a \texorpdfstring{$c$}{c}-structure \texorpdfstring{$(M,\alpha)$}{(M,alpha)}}
To give $\alpha = (\fden{f}^M : \den{a_1}\X\ooo\X\den{a_n}\to \den{c})_{f : a_1,\ooo,a_n\to c \in \Sig_c}$
is to give terms 
$
\jud {x_1:\den{a_1},\ooo,x_n:\den{a_n}} {e_f} b 
$
for all $f : a_1,\ooo,a_n\to c \in \Sig_c$
such that $\fden{f}^M = \eclas{e_f}$.
Note that $\den{c}=b$.
We represent \alpha as a tuple of terms $e_f$
according to function symbols in $\Sig_c$ 
by the order of datatype constructors
listed in a \code{ctype} declaration of $c$.

\subsubsection*{Axiomatising \texorpdfstring{\fold}{fold}}
We next axiomatise the fold operation as a second-order algebraic theory. 
The type of fold may be chosen as
\begin{equation}\label{eq:formal-fold}
\fold^c_b : (\den{\vec{a_1}}\to b), \ooo,(\den{\vec{a_m}}\to b), (\vec{c}\to c)
\to
(\den{\vec{c}}\to b),
\end{equation}
where the first $m$-arguments correspond to the $c$-structure \alpha.
But in second-order algebraic theory, the codomain of function symbol
\W{must be} a sequence of base types (\Sec \ref{sec:soa}),
so the codomain $(\den{\vec{c}}\to b)$ is inappropriate.
We resolve this by simply uncurrying.
We axiomatise the fold as the function symbol
of the type %\y{-1em}
$$
\fold^c_b : (\den{\vec{a_1}}\to b),\; \ooo,(\den{\vec{a_m}}\to b),(\vec{c} \to c),
\den{\vec{c}} \rTo b.
$$
and we will write a term of it using the notation
$$
\;\foldW c b{\,\vec{x_1}.e_1,\ooo,\vec{x_m}.e_m,\,\tju {\vec y} t}{\vec y}
$$
where each $e_i$ corresponds to 
$\fden{f_i}^M$ for
$f_i\in\Sig_c$ in \alpha.
The last two arguments $\tju {\vec y} t$ and  ${\vec y}$
may  need explanation. The first ``$\tju{\vec y} t$'' describes 
a term with variable binding ``${\vec y}.$'' of types $\vec c$, 
but the second ``$\vec y$'' are free variables of types 
$\den{\vec c}$, which are
different from the first bound variables, but have the same length.
We may write simply 
$$
\;\foldD c b{\,\vec{x_1}.e_1,\ooo,\vec{x_m}.e_m,\;t\,}
$$
by omitting the parameters after ``$;$'', when $|\vec y|=0$.
This is a formalisation of  the mathematical expression 
$\mfold^c_b(\alpha)(\eclas{\jud {\vec{y:c}} t c})$ at the level of semantics
as a syntactic term.

In general, we can consider $\fold^{\vec c}_{\vec b}$
as the fold from ${\vec c}$ to ${\vec b}$.
Hereafter, we assume that any signature \Sig is divided into 
the default signature \SigGen, a \Hi{signature for constructors} \SigCon, and \fold's:
\[
\Sig = \SigGen \union \SigCon \union 
  \set{\fold^{\vec c}_{\vec b} \| {\vec c},{\vec b}\in\BB}.
\]
In other words, we assume that any function symbol other than a default constructor or \fold
is an element of $\SigCon$.

In Fig. \ref{fig:fold}, we give the axioms \FOLD, which axiomatise  \fold
by using the characterisation (\ref{eq:fold-law})
in case of a particular category $\CC=\TE$ and a $c$-structure
as a second-order algebraic theory.
To ease understanding, we explicitly describe an instance of (5) as
(5') for the case $c=\code{CList},\, d=(\coSym)$.
Note that there is no case of $\srecEcb{\tju {\vec y} z;\vec y)}$ for $z \not\in\set{\vec y}$.

\begin{rulefigh}
\yy{-2em}
\[%\small
\renewcommand{\arraystretch}{1.4}
\arraycolsep = .3mm
\begin{array}[h]{lllllrllllllll}
(1)&\srecEcb{\tju {\vec{y^c}} {y_i^c}} \parab &=&  {y_i}^b \qquad 
(\text{for }y_i^c \in\set{    {y_1^{c_1}},\ooo,    {y_n^{c_n}}  }  )\\
(2)&\srecE c {\emptytype}{\tju {\vec y} \EMP} \para &=&  \EMP \\
(3)&\srecE {\vec{c_1},\vec{c_2}} {\vec{b_1},\vec{b_1}}{\tju {\vec y} {<\va s[{\vec y}], \va t[{\vec y}]>}} \para &=& 
  {<\foldP {\vec{c_1}} {\vec{b_1}} {\vec y} {\va s[{\vec y}]}  ,\; 
    \foldP {\vec{c_1}} {\vec{b_2}} {\vec y} {\va t[{\vec y}]}  >} \\
(4)&\srecE{\vec c}{\vec b}{\tju {\vec y} \cy{{\vec x}.\va t[{\vec y},{\vec x}]}} \para &=& 
  \cy{{\vec x}. 
       \srecE{\vec c}{\vec b}{\tju {{\vec y},{\vec x}} \va t[{\vec y},{\vec x}]};\,\vec y,\vec x)}
\\
(5)&
\multicolumn{4}{l}{
\srecE{c}{b}{\tju {\vec y} d(\vec{\va a},\va t_1[\vec y],\ooo,\va t_n[\vec y])}
;\vec y ) = 
  ({\vec x. {\va e_d[\vec{\va a},\vec x]}) \at\; 
   <\foldP{c_1}{b_1}{\vec y} {\va t_1[{\vec y}]} ,\ooo
   >
}}
\\
(5')&\srecE {\List}b {\tju {\vec y} {\va a \co {\va t[{\vec y}]}}} \para &=& 
  ({ x. {\va e[\va a, x]}) \at\; 
   \foldP \List b {\vec y} {\va t[{\vec y}]})  }
\\
(6)&\srecEcb{\tju {\vec y} {({\vec x}.\va t[{\vec x}])\at \va s[{\vec y}] }} 
\para &=&  
   (\vec x.\srecEcb{\tju {\vec x} {\va t[{\vec x}]}};\vec x))\; \at\; 
           \srecExx{\vec a}{\vec{a'}}{\tju {\vec y} {\va s[{\vec y}]}  ;\;{\vec y}}
\end{array}
\]

\medskip
\begin{minipage}[t]{\textwidth}\small
Here $E$ is a sequence $(\vec z,\vec x.\va e_d[\vec z,\vec x])_{d\in\Sig_c}$ of metavariables and
$d\in\Sig_c$. (5') is an instance of (5) for explanation.
In (8), $\btt$ and $\bss$ are short for $\va t[\vec x,\vec y]$ and $\va s[\vec x,\vec y]$,
respectively.\\
\end{minipage}
\y{-1.5em}
\caption{Second-order algebraic theory \AxFOLD}
\label{fig:fold}
\medskip
\hrule height 0.5pt
\medskip

\[%\small
\renewcommand{\arraystretch}{1.4}
\arraycolsep = .2mm
\begin{array}[h]{lllllrllllllll}
\ritm 1 \; &\srecEcb{\binder {\vec{y}} {\ovar{y_i}}} \para &\to& 
    {\ovar{y_i}} \\
\ritm 2 &\srecE c {\emptytype}{\binder {\vec y} \EMP} \para &\to& \EMP \\
\ritm 3 &\srecE {\vec{c_1},\vec{c_2}} {\vec{b_1},\vec{b_1}}{\tju {\vec y} {<\va s[{\vec y}], \va t[{\vec y}]>}} \para &\to& 
  {<\foldP {\vec{c_1}} {\vec{b_1}} {\vec y} {\va s[{\vec y}]}  ,\; 
    \foldP {\vec{c_1}} {\vec{b_2}} {\vec y} {\va t[{\vec y}]}  >} \\
\ritm 4&\srecE{\vec c}{\vec b}{\tju {\vec y} \cy{{\vec x}.\va t[{\vec y},{\vec x}]}} \para &\to& 
  \cy{{\vec x}. 
       \srecE{\vec c}{\vec b}{\tju {{\vec y},{\vec x}} \va t[{\vec y},{\vec x}]};\,\vec y,\vec x)}
\\
\ritm 5 & 
\srecEcb{\tju {\vec y} d(\vec{\va a},\va t_1[\vec y],\ooo,\va t_n[\vec y])}
&;\vec y ) &\to&
  {\va e_d[\vec{\va a},
  \foldP{c_1}{b_1} {\vec y} {\va t_1[{\vec y}]} ,\ooo}\\
&&&& \x{2.8em}   
 %\foldP c b {\vec y} {\va t_n[{\vec y}]}
  \foldP{c_n}{b_n} {\vec y} {\va t_n[{\vec y}]}
   ]
\\
\axtitx{Composition}
\\
\ritm{7} &(\binder {\vec y} {\va t[\vec y]})\at <\vec{\va s}> &&\to& 
\va t[\vec s]
\\
\end{array}
\]
\y{-1em}
\caption{Second-order rewrite system \FOLDr}
\label{fig:FOLDr}

\medskip
\hrule height 0.5pt
\medskip

\[%\small
\renewcommand{\arraystretch}{1.4}
\arraycolsep = .3mm
\begin{array}[h]{lllllrllllllll}
\axtitx{\Bekic}
\\
\ritm{10}\;\; &\cyN {m+n}{\vec x,\!\vec y\!.\,<\,\btt,\; \bss\,>} &&\to&
<\, \cyN m{\vec x.\, (\vec y.\btt) \at    \cyN n{\vec y. \bss}     }, \\ %\;\,
&&&& \ \cyN n{\vec y.\, (\vec x.\bss) \at  \cyN m{\vec x.\, (\vec y.\btt) \at  
\cyN n{\vec y. \bss}}}>
\\
\multicolumn{5}{l}{
\textbf{Cleaning rules for $c$ satisfying $\AxBr(\vnil,+)$}
}
\\
\ritm{11} &\cy{x^\Varc.\Uni{\vv x}{\va t}}&&\to& \va t
\\
\ritm{12} &\cy{x^\Varc.\Uni{\va t}{\vv x}}&&\to& \va t
\\
\ritm{13} &\cy{\vec y\!.\va t} &&\to&\va t
\\
\ritm{14} &\cy{x^\Varc.\ovar x} &&\to& \vnil
\\
\ritm{15} &\vnil + \va t &&\to&\va t
\\
\ritm{16} &\va t + \vnil&&\to&\va t
\end{array}
\]

\begin{minipage}[t]{\textwidth}\small
In (7r), $|\vec{\va s}|=1$ is also allowed and
the part $<\vec {\va s}>$ is simply a single metavariable $\va s$.\\
By the notation of metavariable, in (13r),
$\cy{\vec y\!.\va t}$  means that $\va t$ cannot involve $\vec y$, and
similar for (12r)(13r).
The rules (11r)-(16r) are parameterised by the function symbols $\vnil,+$ 
as in $\AxBr(\vnil,+)$.
\end{minipage}
\caption{Second-order rewrite system \SIMP for simplification}
\label{fig:SIMP}

\end{rulefigh}

We omit writing contexts and types of equations
for simplicity. %, which may be clear from terms.
For example, formally the axiom (1) in \AxFOLD is written as
\[
\mju {
 \va e_{1} : \vec{a_1}\to c,\ooo, \va e_{n} : \vec{a_1} \to c
}{}{
\foldW c b {\va e_{1},\ooo, \va e_{n},\,{\tju {\vec y} {y_i}}} {\vec y} 
\;\;=\;\; y_i
}{\quad b}
\]

The arguments of \fold
expressing the $c$-structure are abbreviated as $E$ for simplicity. 

We state the correctness of this axiomatisation, which holds 
by the above faithful construction.
The following is immediate by construction.

\Prop[th:fold-faith]
The following are equivalent.
\begin{itemize}
\item $\mfold^c_b(\alpha)(\eclas {\jud \Gamma t c}) = \eclas{\jud {\Gamma'} u
     b}$ 

\item $\jud {}{\foldW c b {\vec{x_1}.e_1,\ooo,\vec{x_m}.e_m,\; \tju {\vec 
        y}t} 
{\vec y} =  u} {b}$
is derived from the axioms $\EE \union \AxFOLD$ using the second-order equational logic.
\end{itemize}
where \alpha, $e_i$ and $t$ are \fold free, 
$\Gamma=y_1:c,\ooo,y_n:c,\; \Gamma'=y_1:\den{c},\ooo,y_n:\den{c}$.
\oProp

\Example[ex:plus]
The plus function on \code{CNat} can be defined as fold as follows.

\begin{CVerbatim}[commandchars=\\\{\},codes=\mathcom]
  plus : CNat,CNat \tto CNat
  spec plus(m, n) = pl(m)
    where pl(0)    = n
          pl(S(m)) = S(pl(m))
  fun plus(m, n) = fold (n, x.S(x)) m
\end{CVerbatim}

We specify \code{plus} in terms of a unary function
\code{pl} which recurses on the first argument $m$ and gives the second
argument $n$ if $m=0$. Hence it is defined by \fold
where the target \code{CNat}-structure is defined to be 
$\den{\code 0}^M=n, \den{\code S}^M(x)=\code S(x)$.
\oExample

\subsection{Primitive recursion by fold}\label{sec:prim}

The fold axiomatised above covers the ordinary fold on algebraic datatypes.
Thus, we expect that 
various techniques on fold developed in functional programming, 
such as the fold fusion technique
and representation of recursion principles 
such as \cite{DBLP:conf/fpca/MeijerFP91} 
may be transferred to the current setting.
Here we consider a way to implement 
a particular pattern of recursion appearing often in
specifications as a fold.

\begin{framed}
\subsec{Assumptions}%\bigskip
\noindent
Let $(c,\Sig_c,\EE_c)$ and
$(b,\Sig_b,\EE_b)$ be datatype declarations,
where $\Sig_c=\set{d_1,\ooo,d_n}$.
We suppose the following form of specification and definition.
\begin{equation}
\arraycolsep = 1mm
\begin{array}[h]{lllcll}
  &f& \quad : \quad c &\to& b\\ 
  &\texttt{spec }&f(d_1(\vec a,\vec t)) &=& e_{d_1}\\
   &             &&\ccc&\\
    &            &f(d_n(\vec a,\vec t)) &=& e_{d_n}\\
\\
  &\texttt{fun }& \multicolumn{3}{l}{
f(t) = \pi_1 \at \fold^c_{b,c}
  (\, ({v_1,w_1,\ooo,v_n,w_n}.<e'_d,\, d(\vec a,\vec w)>)_{d\in \Sig_c},\; t\,)
}
 \\
\end{array}
  \label{eq:diamond}
\end{equation}
We define $\pi_1 \deq ({v,w}.\; v)$.
We assume that every $e_d$ is a closed term of type $b$,
which may involve terms of 
the types
\[
 \ju{f(t_i)}{ }{b} \qquad \;\ju{t_i}{ }{c}
\]
for each $i = 1,\ooo, n$,
(i.e., $\vec t = t_1,\ooo,t_n$ may appear solely, cf. examples in \Sec
\ref{sec:prog} in $e_d$).
We define $e'_d$ to be a term obtained from $e_d$, by
replacing every $f(t_i)$ with 
$v_i$ of type $b$,
and every $t_i$ (not in the form $f(t_i)$) with $w_i$ of type $c$,
and assume 
$$\;\ju{e'_d}{ v_1\: b, w_1\: c,\ooo, v_n\: b, w_n\: c}{b}
.\;$$
\end{framed}

The above specification can be seen as describing 
primitive recursion, because it is similar to the primitive recursion on natural numbers $f(S(n))=e(f(n),n)$,
where both $n$ and $f(n)$ can be used at the right-hand side.
In functional programming, it is known that 
primitive recursion on algebraic datatypes can be represented as 
fold, called paramorphism \cite{para}.
We now take the fold where the target \Sig-structure is 
the sequence $b,c$ of types,
i.e., $\fold^c_{b,c}$.

\subsec{Structure}
Let $\EE$ be the set of axioms collecting the axioms of all datatypes in \BB.
Hence $\EE \supseteq \EE_b \union \EE_c$.
We define a $(\Sig_c,\EE)$-structure $(M,\alpha)$ in \TE by
\[
  \den{c}^M  = (b,c)
\]
and $\den{a}^M = a $, otherwise.
If $\EE_c$ contains axioms $\nAxBr(\vnil,+)$, then
$\EE_b$ must contains axioms $\nAxBr(\nu,\mu)$,
where $\nu : b,\; \mu:b,b\to b$ are  in $\Sig_b$.
We define
\[
\arraycolsep = 1mm
\begin{array}[h]{llrllllllllllllllll}
\fden{\, \vnil \,}^M &=& <\unit,\vnil>,\qquad
\fden{\uniSym}^M(<v_1,w_1>,<v_2,w_2>) &=& <\mu(v_1,v_2),w_1+w_2>
\\
&& \fden{d}^M(\vec a, v_1,w_1,\ooo,v_n,w_n) &=& <e'_d,\, d(\vec a,\vec w)>
\end{array}
\]
for $d : \vec a,c^n \to c \in \SigCon$.

\subsec{Formalisation}
Then we have the function 
\begin{equation}
\mfold^c_{b,c}(\alpha) : \TE(\vec{c},\; c)\rTo\TE(\den{\vec{c}},\; (b,c))
\end{equation}
which is axiomatised as the function symbol
$$
\fold^c_{b,c} : (\den{\vec{a_1}}\to {(b,c)}),\; \ooo,(\den{\vec{a_m}}\to 
{(b,c)}),(\vec{c} \to c),
\den{\vec{c}} \rTo_{} (b,c).
$$
The axioms \FOLD in Fig. {\ref{fig:fold}} is instantiated to the case of
$\fold^c_{b,c}$. For example,
\[%\small
\renewcommand{\arraystretch}{1.3}
\arraycolsep = 1mm
\begin{array}[h]{lllllrllllllll}
(1)\qquad&\srecE c {b,c}{\tju {\vec{x}} { x_i }} &;\;v_1,w_1,\ooo,v_n,w_n) &=& <v_i,w_i>
\end{array}
\]
Other axioms for \fold are instantiated similarly.

\subsec{Properties}
By construction, 
$$\ju{e_d \;\,=\;\, 
(\tju {v_1,w_1,\ooo,v_n,w_n}{e'_d}) \at <f(t_1),t_1,\ooo,f(t_n),t_n>}{ }{b}   
\label{eq:eabs}
$$
holds.
Define $E \deq ({v_1,w_1,\ooo,v_n,w_n}.{e'_d})_{d\in \Sig_c}$.
By induction on the typing derivations, we have 
$$\fold^c_{b,c}(E ,\; t) 
 = {<\, f(t),\;  t  \,>}$$
for all closed terms $t$ of type $c$.
By the characterisation (\ref{eq:fold-law}), we have 
for each $d\in\SigCon$,
\begin{meqa}
<f(d(\vec a,\vec t)),\, t>
&= \fold^c_{b,c}(E ,\; f(d(\vec a,\vec t))) \\
&= <({v_1,w_1,\ooo,v_n,w_n}  . <e'_d,d(\vec a, \vec w)> )
  \at <\, f(t_1),t_1,\ooo,f(t_n),t_n\, >,\, t> = <e_d,\,t>
\end{meqa}
hence we have 
\[
f(d(\vec a,\vec t)) = e_d
\]
meaning that $f$ satisfies 
the specification.
We use this representation of primitive recursion in \Sec \ref{sec:prog}.

\Remark
If $e_d$ constrains only recursive calls of the forms $f(t_i)$, it is 
merely a pattern of structural recursion, so
it can be implemented by \fold using the structure $v.e'$ where 
all the recursive calls $f(t_i)$ in $e_d$ are abstracted to $v$ as
Example \ref{ex:plus}.
\oRemark

% Local Variables:
% TeX-master: "fs"
% End:

\section{Strongly Normalising Computation Rules for \FOLD}\label{sec:SN}
We expect that $\AxFOLD$ provides
strongly normalising computation rules.
An immediate idea is to regard the axioms \AxFOLD as rewrite rules by orienting 
each axiom from left to right.

But proving strong normalisation (SN) of \AxFOLD is not straightforward.
The sizes of both sides of equations in \AxFOLD are not decreasing 
in most axioms.
So, assigning some ``measure'' to the rules in \FOLD
that is strictly decreasing is difficult for this case.
Such a naive method is typically dangerous for higher-order rewrite rules
and might lead one to an unintentional mistake. %, such as in \cite{Ohta-correct}.
If the axioms regarded as second-order rules are a binding 
Combinatory Reduction System (CRS) \cite{CRS} (cf. Remark \ref{rem:metaterms}),
meaning that every {meta-application}
$\m{\va{m}}{t_1,\ooo,t_n}$ is of the form $\va m [\vec x]$,
then it is possible to use a simple
polynomial interpretation to prove termination of second-order rules 
\cite{CRS}.
Unfortunately, this
is not the case because in (5)
there is a meta-application 
violating the condition.
Existence of meta-application means that it essentially
involves the \beta-reduction,
thus it has the same difficulty as proving
strong normalisation of the simply-typed \lmd-calculus.

We use a general established method of \W{\GS} \cite{IDTS,IDTS00},
which is based on Tait's computability method to show \W{strong
  normalisation} (SN).
\TGS has succeeded to prove SN of various recursors such as 
the recursor in G\"{o}del's System T.
The basic idea of \GS is to check whether the arguments of 
recursive calls in the right-hand side of a rewrite rule 
are ``smaller'' than the left-hand sides' ones. It is similar 
to Coquand's notion of 
``structurally smaller'' \cite{CoqPat}, but more relaxed and extended.

\subsection{General Schema}

We review \GS criterion \cite{IDTS00,Blanqui-TCS}.
For more details and the proofs, see the original papers
\cite{IDTS00,Blanqui-TCS}.

\TGS %in \cite{IDTS00} 
is formulated for 
a framework of %second-order 
rewrite rules called 
inductive datatype systems, whose second-order fragment
is essentially the same as 
the present formulation given in \Sec \ref{sec:salg}.
Minor differences are adapted as follows.
We use greek letters such as $\alpha,\sig,\tau,\ooo$  to
denote types. Roman alphabets such as $a,b,c,\ooo$ are used
for base types.
\begin{enumerate}
\item The target of function symbols must be a single
(not necessary base) type in an inductive datatype system.
Hence we introduce the product type constructor 
$\X$, assume that $b_1 \X b_2$ is again a base type
in the sense of \Sec \ref{sec:cytype}, and use it for
the target type. 

\item Instead of a term $x_1,\ooo,x_n.t$ of sort $a_1,\ooo,a_n\to b$ in
  a second-order algebraic theory,
we use $x_1.\ccc.x_n.t$ of type $a_1\to\ccc\to a_n\to b$.
Now the abbreviation $\vec x.t$ denotes $x_1.\ccc.x_n.t$.

\end{enumerate}

\Def
A set of rewrite rules induces the following relation on function symbols
in a signature \Sig:
$f$ \W{depends on} $g$ if there is a rewrite rule defining $f$ 
(i.e., whose left-hand side is headed by $f$) in the right-hand side of
which $g$ occurs.
Its transitive closure is denoted by $\gt_\Sig$.
\oDef

\Def[def:constr]
A \W{constructor} is a function symbol $c : \vec \tau \to b$ 
which does not occur at the root symbol of the left-hand side 
of any  rule.
The set of all constructors 
defines a preorder $\le_\BB$ on the set \BB of mol types by 
$a \le_\BB b$ if 
$b$ occurs in $\vec \tau$
for a constructor $c : \vec \tau \to b$.
Let $\lt_\BB$ be the strict part and $=_\BB$ the equivalence relation
generated by $\le_\BB$.
A type $b$ is \W{positive} if 
for any type $a$ s.t. $a =_\BB b$, $a$ does not occur at a 
negative position 
of the type of constructor $c$ of $b$.
A constructor $c :\vec\tau \to b$ is positive 
if $b$ is positive.
\oDef

\Def[def:acc]
A metavariable $\va z$ is \W{accessible} in a meta-term $t$ if
there are distinct bound variables $\vec x$ such that $\m{\va{z}}{\vec x}
\in \Acc(t)$, where $\Acc(t)$ is the least set satisfying
the following clauses:
\begin{enumerate}[leftmargin=*,label={{(a\arabic*)}}]
\item $t \in \Acc(t)$.
\item If $\q{x}{u} \in \Acc(t)$ then $u \in \Acc(t)$. 
\item If $c(\vec{s})\in\Acc(t)$ then each $s_i \in \Acc(t)$
for a constructor $c$.
\item  Let $u_i$ be a term of type $\tau_i$ for each $i=1,\ooo,n$.
If $\Acc(t) \ni f(u_1,\ooo,u_n)$ is of a base type $b$,
and $b$ (possibly) occurs positively in type $\tau_i$, then $u_i\in \Acc(t)$
\cite[Def.15]{Blanqui-TCS}.

\end{enumerate}
\oDef

\noindent A position $p$ is a finite sequence of natural numbers.
The order on positions is defined by $p \lt q$ if there exists
a non-empty $p'$ such that $p\, p' = q$.
Given a meta-term $t$, 
$\Pos(t)$ denotes the set of all positions of $t$.
The notation $\subterm{t}p$ denotes a subterm of $t$ at a position $p$, and
$\replace{t}{s}p$
the term obtained from $t$ by replacing the subterm at the
position $p$ by $s$.

\Def
A meta-term $u$ is a
\W{covered-subterm} of $t$, written $t \cge u$,
if there are two positions $p\in \Pos(t), q \in \Pos(\subterm{t}p)$ 
such that 
\begin{itemize}
\item $u = \replace t{ \subterm{t}{pq} }p$,
\item $\forall r \lt p.\; \subterm{t}r$ is headed by an abstraction, and
\item $\forall r \lt q.\; \subterm{t}{pr}$ is headed by a function symbol,
including a constructor.
\end{itemize}
\oDef

For example, $f(a,c(x)) \cge c(x)$, and $\lmd(x.\va m[x]) \cge \va m[x]$,
and $y.\lmd(x.\va m[x]) \cge y. x.\va m[x]$.

\Def[def:ccl]
Given $f : \tau_1,\ooo,\tau_n \to \tau \in \Sig$,
the \W{computable closure} $\CCl_f(\vec t)$ of a meta-term $f(\vec t)$ is
the least set \CCl satisfying the following clauses.
We assume that all the terms below are well-typed.
\begin{enumerate}[leftmargin=*,label={{(\arabic*)}}]
\item If $\va{z} : \tau_1,\ooo,\tau_p \to \tau$ is accessible in 
some of $\vec t$, and $\vec u \in \CCl$,
then $\va z [\vec u]\in \CCl$.

\item For any variable $x$, $x\in \CCl$.

\item If $c : \tau_1,\ooo,\tau_n \to b$ is a constructor
and $\vec u \in \CCl$,
then $c(\vec u)\in\CCl$.

\item If $u,v \in \CCl$, then $u@v\in\CCl$ for $@: (\sig\to\tau),\sig\to\tau$.

\item If $u\in \CCl$ then $x.u\in \CCl$.

\item If $f \gt_\Sig g$ and $\vec w\in \CCl$, then
$g(\vec w)\in \CCl$.

\item If $\vec u\in \CCl$ such that 
$\vec t \cgtmul \vec u$, 
then $f(\vec u)\in \CCl$,
where $\cgtmul$ is the lexicographic extension
of the strict part of $\cge$.
\end{enumerate}
\oDef

\Def
A rewrite rule $f(\vec t)\to r$
\W{satisfies} \GS 
if $\CCl_f(\vec t) \ni r $.
\oDef

\ThTitled[th:GS]{\cite{IDTS00}}
Suppose that given a signature \Sig and
rules \RR satisfies the following:
\begin{enumerate}[noitemsep,label={{(\arabic*)}}]
\item $\gt_\BB$ is well-founded,
\item every constructor is positive, and
\item $\gt_\Sig$ is well-founded.
\end{enumerate}
If all the rules of \RR satisfy \GS, then
$\RR$ 
is strongly normalising.
\oTh

% Local Variables:
% TeX-master: "fs"
% End:

\subsection{Refining second-order algebraic theory to second-order rewrite rules}
In order to apply \GS criterion, %to a rewrite system of \AxFOLD,
we refine the second-order algebraic theories \AxCy, \AxBr and \AxFOLD 
to
the second-order rewrite rules \FOLDr and \SIMP.

Crucially, 
the constructors used in \FOLD are {not positive},
as $\fcy$ and $\at$ involve a negative occurrence of $c$ in $(c\to c)$.
We can overcome this problem by 
modifying the type $(c\to c)$ to a restricted %one
$(\Var_c\to c)$, where $\Var_c$ is 
a base type having no constructor
considered as the type of ``variables'' of type $c$. 
We assume the constructor $\fn{v}$ which embeds a ``variable''
into a term.
We %now 
modify the types of default constructors as follows:
\[
\begin{array}[h]{cllcllll}
<-,\ccc,-> &: {c_1},\ooo,{c_n}\to {c_1}\X\ooo\X{c_n},
&&\cyc &:(\vec{\Var_{c}}\to c)\to c,\\
\fn{v} &: \Var_c \to c,
&&%-\at - &: (a_1,\ooo,a_n\to c),a_1\X\ccc\X a_n \to c,
\end{array}
\]
where $c$'s and $a$'s are base types of the inductive datatype system
(which may be product types), $\vec{\Var_{a}}\to c$ is
short for $\Var_{a_1}\to\ccc\to\Var_{a_n}\to c$.
The type of \fold is now
$$
\fold^c_b : (\vec{\Var_{a_1}}\to b), \ooo,(\vec{\Var_{a_k}}\to b), 
(\Var_{c}^m\to c), \Var_{b}^m \rTo b,
$$
The use of a type $\Var_\sig \to \tau$ to represent binders 
is well-known in the field of mechanized reasoning, sometimes called
\W{(weak) higher-order abstract syntax} \cite{HOAS-Coq}.

We also modifty the type of composition as
\[
-\at - : (a_1,\ooo,a_n\to c),a_1\X\ccc\X a_n \to c
\]
which is not a constructor in the sense of Def. \ref{def:constr},
hence it is allowed as not positive (when $a_i=c$).

\subsection{Second-order rewrite systems \FOLDr and \SIMP}\label{sec:apd}

We now describe how we obtain 
the second-order rewrite rules \FOLDr and \SIMP
given in Fig. \ref{fig:FOLDr} and \ref{fig:SIMP}.
Note that \FOLDr's ``\textsf{r}'' stands for ``rewrite''.
The rewrite system \FOLDr is an oriented version of \FOLD.
The axiom (5) is refined to (5r) by applying \textbf{Composition} rules
which is need to prove SN.
Because of it, a rule corresponding to (6) is not needed in \FOLDr.
The axiom (sub) in \AxCy involves a meta-application at the right-hand side, 
which performs
general substitution of terms for variables $\vec y$ of base types.
The oriented version of it is the rule (7r).

The rewrite system \SIMP is for simplification, whose rules
are taken from the 
equational theory of $\AxCy\union\AxBr$.
We include the \Bekic law as a rewrite rule (10r),
which can be depicted as:
\begin{center}
\includegraphics[scale=.32]{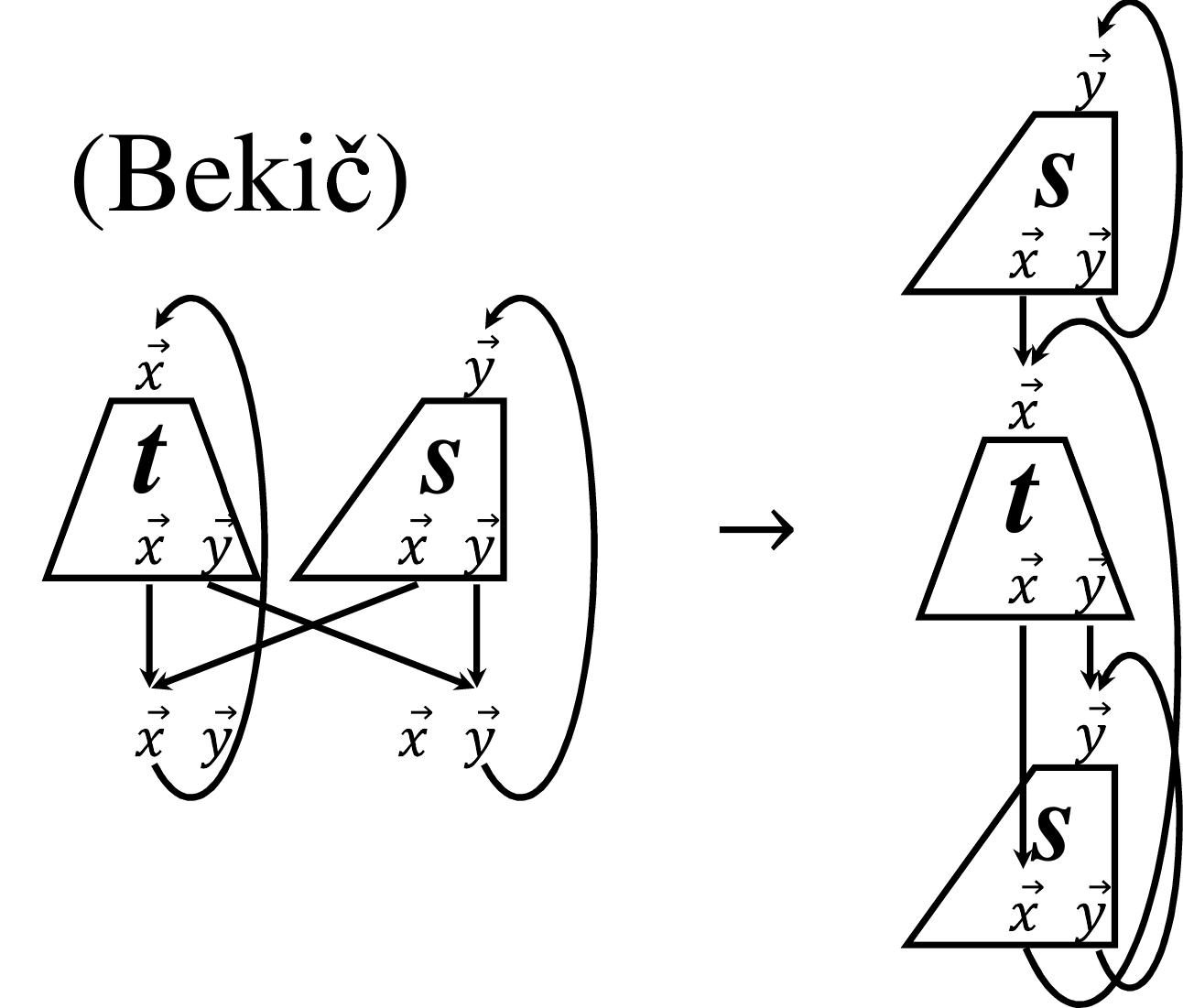}  
\end{center}
It says that the fixed point of a pair 
can be obtained by computing the fixed points of its components
independently and composing them suitably (see the right figure).
It can be seen as \Hi{decreasing} complexity of cyclic computation
because looking at the argument of \fcyc,
the number of components of tuple is reduced.
The superscript of $\fcy$ in (10r)
indicates the length of the tuple argument (see \Sec \ref{sec:soa}).
Such a superscript indicating an invariant of the arguments
is similar to the idea of higher-order semantic labelling
\cite{HSL},
but here we just make the existing superscript explicit rather than labelling.
Hence it is suitable for rewrite rules and actually shown to be terminating 
(cf. the proof of Thm. \ref{th:fold-sn}).

We define the relation $\to_\FOLDr$ (resp. $\to_\SIMP$)
on terms by 
the relation generated by 
the second-order equational logic Fig. \ref{fig:sel}
under the axioms \FOLDr (resp. \SIMP)
without using the (Ax2), (Ref) and (Tra)-rules.
Namely, ``one-step rewriting'' $\to_\FOLDr$ 
is equational reasoning without using
symmetry, reflexivity and transitivity.
By construction, we immediately see that
the rewrite system \FOLDr correctly implements the second-order algebraic
theory \FOLD.
The following proposition is immediate by construction of rules.
We write $\check t$ for a term that
recovers the original terms $t$ by stripping 
the constructor $\fn{v}$.

\Prop% \\[-1em]
If $t \to^+_\FOLDr t'$,
then $\check t = \check{t'}$ is derivable from $\FOLD\union \AxCy\union\AxBr$.
\oProp

\bigskip
\Th[th:fold-sn]
The second-order rewrite system $\FOLDr\union\SIMP$ is strongly normalising.
\oTh
\proof We use the \GS criterion
using the well-founded relation $\gt_\Sig$  on function symbols
$$
\fold,\at \gt \cyc^m \gt \cyc^n\gt \text{ any other constructors}
$$
where natural numbers $m \gt n \ge 1$.
Now all constructors are positive.

In the following proof, we write $f(\vec t) \gtt s\;$ when $\;\CCl_f(\vec t) \ni s$.
We show that $\FOLDr\union\SIMP$ satisfies \GS,
i.e., for each rewrite rule $f(\vec t) \to s$, we check $f(\vec t) \gtt s$.

\begin{itemize}[leftmargin=2em]
\item [(1r)]
$\fold(E,\vec y.\vv{y_i};\vec y) \gtt \vv{y_i}
\by{6} \fold(E,\vec y. \vv{y_i};\vec y) \gtt y_i,$
which holds by (2).
\end{itemize}

\noindent
The implication ``$\Longleftarrow$'' expresses backward inference, which
is labelled with a number indicating
which clause in Def. \ref{def:acc} or Def. \ref{def:ccl}
is applied.
In what follows, for brevity, we examine mostly the cases that the length of bound variables 
is $1$ (i.e., $|\vec x|=|\vec y|=1$) and the arity of $d$ is $2$.
General cases are proved similarly. 

\begin{enumerate}[label=(\alph* r),widest=10]

\item [(4r)]
$\fold(y.\cyc(x.\va t[y,x]);y) \gtt \cyc(x.\fold(y,x.\va t[y,x];y))
\\\by{6} \fold(y.\cyc(x.\va t[y,x]);y) \gtt \fold(y,x.\va t[y,x];y)
\\\by{7} 
         \fold(y.\cyc(x.\va t[y,x]);y) \gtt y
\\\ANDD
         \fold(y.\cyc(x.\va t[y,x]);y) \gtt y,x.\va t[y,x]
\ANDD
y.\cyc(x.\va t[y,x]) \cge y,x.\va t[y,x]
$

\bigskip\noindent
The second literal holds easily.
In general, if a metavariable with binders is a subterm of a larger meta-term,
it is proved to be smaller by $\gtt$.
The third literal of a covered subterm relation holds, because stripping the prefix binders,
it is a subterm relation.

\item [(5r)]
$\fold(z,x.\va e_d[z,x],y.d(\va a,\va t[y]);y) \gtt 
\va e_d[\va a, \fold(z,x.\va e_d[z,x],y.\va t[y];y) ]  
\\\by{6} \fold(z,x.\va e_d[z,x],y.d(\va a,\va t[y]);y) \gtt \va a,\; 
 \fold(z,x.\va e_d[z,x],y.\va t[y];y)\\ \quad \ANDD \quad \va e_d
$
is accessible in $(z,x.\va e_d[z,x])$.

\medskip

\noindent
The first literal is proved to be smaller straightforwardly (cf. (4r))
and the accessiblity 
$
\Acc(z,x.\va e_d[z,x]) \ni \va e_d[z,x]
$
is shown by
(a1)(a2).
The general case that the part $x.\va e_d[x]$ in \fold
is a sequence of metavariables
$(x.\va e_d[x])_{d\in\Sig_c}$ is similar.
Hereafter, we omit writing the part $E=(x.\va e_d[x])_{d\in\Sig_c}$ 
and the arguments after ``;'' 
in \fold for brevity.

\item [(3r)]
$\fold(y.<\va s[y],\va t[y]>) \gtt <\fold(y.\va s[y]),\fold(y.\va t[y])>
\\ \by{3} \fold(y.<\va s[y],\va t[y]>) \gtt \fold(y.\va s[y]) \ANDD 
     \fold(y.<\va s[y],\va t[y]>) \gtt \fold(y.\va t[y])
\\\by{7}  y.<\va s[y],\va t[y]> \cge y.\va s[y]  \ANDD  y.<\va s[y],\va t[y]> \cge y.\va t[y]
$

\item [(7r)]
$(\binder {\vec y} {\va t[\vec y]})\at <\vec{\va s}> \gtt \va t[\vec{\va s}]
\\\by{1}
(\binder {\vec y} {\va t[\vec y]})\at <\vec{\va s}> \gtt \vec{\va s} \ANDD
\Acc({\vec y.\va t[\vec y]})\ni \va t[\vec y]
$

\medskip
\noindent
The first literal holds because $\vec{\va s}$ are accesible in $<\vec{\va s}>$.
The second literal holds by (a2)(a1).

\item [(10r)]
We write $\btt=\va t[\vec{x},\vec{y}], \bss=\va s[\vec{x},\vec{y}]$.\\
$
\cycMN(\vec{x},\vec{y}.<\btt,\bss>) \gtt <\cyc^m(\vec{x}. (\vec{y}.\btt) \at \cyc^n(\vec{y}.\bss)),
                 \cyc^n(\vec{y}.(\vec{x}.\bss) \at \cyc^m(\vec{x}.(\vec{y}.\btt) \at \cyc^n(\vec{y}.\bss)))>
\\\by{6} \cycMN(\vec{x},\vec{y}.<\btt,\bss>) \gtt \cyc(\vec{x}. (\vec{y}.\btt) \at \cyc(\vec{y}.\bss))           
   \\\phantom{.}\; \ANDD \cycMN(\vec{x},\vec{y}.<\btt,\bss>) \gtt \cyc(\vec{y}.(\vec{x}.\bss) \at \cyc(\vec{x}.(\vec{y}.\btt)
   \at \cyc(\vec{y}.\bss))) \quad \deq \text{(A)} \ANDD \text{(B)}.
$
\; In what follows, unlabelled $\cyc$ denotes $\cyc^m$ or $\cyc^n$.
\begin{enumerate}[label=(\Alph*)]
\item 
[(A)] $\cycMN(\vec{x},\vec{y}.<\btt,\bss>) \gtt \cyc(\vec{x}. (\vec{y}.\btt) \at \cyc(\vec{y}.\bss))
\by{6} \cycMN(\vec{x},\vec{y}.<\btt,\bss>) \gtt \vec{x}. (\vec{y}.\btt) \at \cyc(\vec{y}.\bss)   
\by{6} \cycMN(\vec{x},\vec{y}.<\btt,\bss>) \gtt \vec{x}. (\vec{y}.\btt),\;
  \cyc(\vec{y}.\bss)  
$

\item 
[(B)] $\cycMN(\vec{x},\vec{y}.<\btt,\bss>) \gtt \cyc(\vec{y}.(\vec{x}.\bss) \at \cyc(\vec{x}.(\vec{y}.\btt) \at \cyc(\vec{y}.\bss)))
\\\by{6} \cycMN(\vec{x},\vec{y}.<\btt,\bss>) \gtt \vec{y}.(\vec{x}.\bss) \at \cyc(\vec{x}.(\vec{y}.\btt) \at \cyc(\vec{y}.\bss))
\\\by{6} \cycMN(\vec{x},\vec{y}.<\btt,\bss>) \gtt \vec{y}.(\vec{x}.\bss)       
  \ANDD  \cycMN(\vec{x},\vec{y}.<\btt,\bss>) \gtt \cyc(\vec{x}.(\vec{y}.\btt) \at \cyc(\vec{y}.\bss))
$

\bigskip\noindent
The first literal easily holds. The second literal holds by
\medskip
$
\\\by{6} \cycMN(\vec{x},\vec{y}.<\btt,\bss>) \gtt \vec{x}.(\vec{y}.\btt) \at \cyc(\vec{y}.\bss)
\\\by{6} \cycMN(\vec{x},\vec{y}.<\btt,\bss>) \gtt \vec{x}.(\vec{y}.\btt),\quad
\cyc(\vec{y}.\bss)$,\quad each of which holds easily.  
\end{enumerate}
\end{enumerate}                                

\noindent The remaining cases are simpler and similarly proved.
By Thm.~\ref{th:GS},
we have strong normalisation of $\FOLDr\union\SIMP$.
\QED

\y{-1em}
\Remark
This strong normalisation result is very general.
Not only ensuring termination of computation of 
$\;\foldD c b{\,\vec{x_1}.e_1,\ooo,\vec{x_m}.e_m,\;t\,}$
for
\W{closed terms} $\vec e,t$ (which is an expected result by 
the semantics characterisation), 
the result ensures that \W{any term} 
$\;\foldW c b{\,\vec{x_1}.e_1,\ooo,\vec{x_m}.e_m,\,\tju {\vec y} t}{\vec y}$
involving possibly 
\begin{enumerate}
\item multiple \fold's, or even nested \fold's  \label{item:nested} and
\item free variables (i.e., $e$ and $t$ can be \W{open terms}) \label{item:freevar}
\end{enumerate}
is strongly normalising without imposing any reduction strategy.
For \ref{item:nested}, consider the situation that
one defines a function which calls other functions.
Since in this paper, our methodology is that
any function on cyclic datatypes is defined using \fold,
this situation is realised using \fold involving other \fold's. 
\oRemark

% Local Variables:
% TeX-master: "fs"
% End:

\section{Decidability of Equational Theory}
\label{sec:CR}

In this section, we show an important property of our framework,
namely, the decidability of
the equational theory generated by \AxCy, \AxBr and \FOLDr.
This is done by investigating another important rewriting property,
\W{Church-Rosser modulo bisimulation} of \FOLDr.
SN of \FOLDr established in the previous section also plays
an important role to establish Church-Rosser modulo bisimulation.
Note that since the rewrite rules \SIMP is 
merely a subset of an oriented version of theorems derived from
$\AxCy\union\AxBr$,
we do not need to include \SIMP for the decidability of equational theory.

\Notation\label{nt:normlise}
Hereafter, we omit writing the variable term constructor \fn{v} in terms
used in \FOLDr for simplicity.
For example, we will simply write $\cy{\binder x x}$ to mean
$\cy{\binder x \ovar x}$.
We may write simply $\to$ for the rewrite relation $\to_\FOLDr$.
We write $\too$ for the reflexive transitive closure,
$\to^+$ for the transitive closure, and $\ot$ for the converse of $\to$.
We define $\convone \;\deq\; \to \union \ot$.
The notation $\nf t $ denotes a unique normal form of $t$.
We write 
$t \normalise t'\;$ if $\;t \to^* t'$
and $t'$ is a normal form, meaning 
rewriting to a normal form.
\oNotation

\subsection{Church-Rosser modulo \bisim}

An important property for rewriting with equational theory
is Church-Rosser modulo 
equivalence relation \cite{Huet,Terese}.

A relation $\to$ is \Hi{Church-Rosser modulo \bisim}
(\CRsim)
if $\;s \;(\sim \union \convone)^*\; t\;$ implies 
there exist $s',t'$ such that $s\too s' \;\;\&\;\; t\too t' \;\;\&\;\;
s' \bisim t'$.
In diagram,
\begin{equation}
\begin{diagram}[size=2em]
   s &\quad(\sim \union \convone)^*\quad& t \\
\dDashto<* &&\dDashto>* \\
s' &\bisim& t'
\end{diagram}
\label{eq:CRsim}
\end{equation}
\medskip

\Def[th:simAx]
We define \bisim to be the equivalence relation on terms generated by 
$\AxCy \union \AxBr$, i.e., bisimulation.
\oDef

\subsection{Confluence on plain terms}

We consider confluence on terms without any quotient.

\Prop
The relation $\toFOLDr$ is confluent.
\oProp
\proof
Since the only critical pair between (6) and (7r) is joinable,
$\toFOLDr$ is locally confluent.
$\toFOLDr$ is also SN (Thm. \ref{th:fold-sn}).
By Newman's lemma \cite{Huet,Baader},
it is confluent \cite{Klop}.
\QED

Note that this result does not contradict known counterexamples 
to confluence of 
graph rewriting systems \cite{ETGRS},\cite[Example 2.4.3]{Hassei}. 
Counterexamples to confluence in \cite{ETGRS} use 
the fixed point law as a rewrite rule, while our \FOLDr does not have the 
fixed point law.

\subsection{Normal forms}
\Prop
The normal form of a term by rewriting with \FOLDr is unique.
\oProp
\proof 
Since \FOLDr is confluent.
\QED

We analyse the structure of normal forms.
We call a term a \W{value} if it follows the grammar $(d\in \SigCon)$:
\[
  t,t_1,\ooo,t_n ::= {{y}}
  \|   d(\vec t)
  \|   \cy{\vec x.{t}} 
  \|   <t_1,t_2>
  \|   \EMP
  \|   (\vec y.t_1) \at t_2
\]

\Prop[th:value]%\ \\[-1em]
Suppose
$\jud{\vec{x_1}}{e_1}c,\,\ooo,\,\jud{\vec{x_m}}{e_m}c\;$ and
define $e \deq \vec{x_1}.e_1,\ooo,\vec{x_m}.e_m$.
If 
$
\jud {\vec {y:b}} t {\vec c} 
$ is a value,
then
$\nf{\;\foldU{e, \vec y.t}{\vec y}\;}$ is a 
value.

\label{itm:value}

\oProp
\proof
We abbreviate
$\fold(e,\vec y.t; \vec y)$ as 
$\fold(t)$.
By induction on the structure of values.
Base cases: $y$ and $\EMP$'s normal forms are themselves.
Induction step: Case $t=<t_1,t_2>$ is a value.
\[
\nf{\fold(<t_1,t_2>)}
=
\nf{<\fold(t_1),\fold(t_2)>} = <\nf{\fold(t_1)},\nf{\fold(t_2)}> \eqIH <v_1,v_2>
\]
where $v_1,v_2$ are values.
The cases $d(\vec t),\cy{x.{t}},(\vec y.t_1)\at t_2$ are similar.
\QED

The propositions Prop. \ref{th:fold-faith} and Prop. \ref{th:value}
show that
\AxFOLD is a correct implementation of 
the characterisation (\ref{eq:fold-law})  
stating that \fold preserves bisimilarity and is total.

\subsection{A decidable proof method for the equational theory}
\label{rem:CRsim}
\CRsim is a desirable property, because the diagram (\ref{eq:CRsim}) 
describes
a proof method of $\;s \;(\sim \union \convone)^*\; t$,
i.e., $s=t$ is derivable from 
$\AxCy\union\AxBr$ and the equational theory generated form \FOLDr.

We explain the reason why \CRsim is useful below.
We suppose that \CRsim holds.
Now \CRsim means that a proof of $$\;s \;(\sim \union \convone)^*\; t,\;$$
where $s$ and $t$ are connected by disordered combinations of 
$\sim,\;\to,\;\ot,$
is always transformed to a proof
\[
s \;\too\; s'\; \sim\; t' \;\oot\; t
\]
for some $s',t'$.
But this is still not good enough, because it is not clear
how many times we should rewrite $s$ and $t$ to reach 
suitable $s'$ and $t'$.
For the case of \FOLDr, we can further transform it to
an equivalent proof
(cf. Notation \ref{nt:normlise})
\begin{equation}\label{eq:proof-method}
s \;\normalise\; s_0\; \sim\; t_0 \;\normaliserev\; t  
\end{equation}
which means 
that one first normalises the terms $s,t$ 
to the unique normal forms $s_0,t_0$ by \FOLDr and then
compares them by the equality \bisim.
The reasons why this method is possible are
\begin{enumerate}[noitemsep]
\item $\to$ is SN, %and
\item $\to$ has the unique normal form property
\item $\too$ preserves $\sim$ by \CRsim, and
\item the bisimulation $\bisim$ is decidable.
\end{enumerate}
Hence (\ref{eq:proof-method}) gives a decidable proof method for the equational theory
generated by \FOLDr, \AxCy and \AxBr.

\subsection{Establishing \CRsim}
Unfortunately, \CRsim does not hold on unrestricted terms.
For example, letting $e \deq (l,t.\, 2\co t)$, we have
the following situation, which violates \CRsim.
A waved line in the diagram denotes $\bisim$.
\begin{equation}
\vcenter{
\xymatrix@!C=6.5em{
& & 1\co \fold(e,\fcy(x. 1\co \fold(e,x))) \ar[ld]_-{(4r)(5r)}
    \ar@{~}[rd]^(.6){\txxt{\quad(dinat)(sub)}}& &\\
&1\co \fcy(x. 2\co \fold(e,\fold(e,x))) \ar[d]_{\quad\not} 
&\not\bisim& \fcy(x. 1\co \fold(e,x)) \ar[d]_{\quad\not}\\
& &&
}
}
\label{eq:no-LCH}
\end{equation}

Note that the left-hand side rewrite applies (4r)(5r), which pushes
\fold into the constructor \cyc, and changes $1$ to $2$, because of 
the definition of $e$.
A crucial point is that $\fcy(x. 1\co \fold(e,x))$
involves a normal form $\fold(e,x)$, where 
\Hi{$x$ is bound by a binder placed upward}.
There are three possibilities that such a bound variable $x$ appears:
%\begin{center}
\begin{enumerate}[label=\({\alph*}]
\item  $\fcy(\vec y. C[\;\foldNP e {\vec z} x \;])$\label{itm:bound-a}
\item  $(\vec y. C[\;\foldNP e {\vec z} x \;])\at t$\label{itm:bound-b}
\item  $\fold(e,\vec y. C[\;\foldNP e {\vec z} x)\;])$\label{itm:bound-c}
\end{enumerate}
%\end{center}
where 
\Hi{a variable $x$ is bound by one of binders $\vec y$}, 
and $C$ is a context 
that does not bind $x$.
Now $e = \vec{x_1}.e_1,\ooo,\vec{x_m}.e_m$.
The cases \ref{itm:bound-b} and \ref{itm:bound-c} are no problem,
i.e., they do not 
fall into a situation as the diagram (\ref{eq:no-LCH}).
The only problematic case is \ref{itm:bound-a}. 
Hence, we exclude such terms from consideration.

\Def[def:bad-term]
Given a term $t$,
if $\nf t$ has a subterm of the form
$$\fcy(\vec y. C[\;\foldNP e {\vec z} s) \;]),$$ 
where $s$ involves a variable $x$ bound by one of 
binders $\vec y$ of \fcy,
then we call $t$ a \Hi{bad term}.
We define a set of non-bad terms as follows:
\[
\TT \deq \set{t \| \jud \Gamma t \vec c \text{ with $t$ is not bad,
and every first $m$-arguments $e$ of \fold are closed
}}.
\]
The condition that the first $m$-arguments $e$ 
of \fold (cf. \ref{eq:formal-fold})
are closed is to avoid similar problems.

\oDef

\Prop
The set \TT is closed under the rewrite relation $\toFOLDr$  and 
one step application of an axiom in \AxCy and \AxBr.
\oProp
\proof
If $t\in\TT$ then $t\toFOLDr t' \in \TT$ because \FOLDr does not produce 
a bad term if $t$ does not involve a bad term.
Similarly, an application of an axiom in \AxCy and \AxBr
does not produce 
a bad term if $t$ does not involve a bad term.
Note that substitution of terms for variables are always capture-avoiding.
\QED

Note that a bad term does not simply mean that 
\fold does not appear under \fcy. 
A term where \fold appears under \fcy is allowed, 
when $s$ does not involve the bound variables $\vec y$ of \fcy.
For example, the right-hand side of the rule (4r) in \FOLDr
is not bad.

We use the following theorem to show \CRsim.

\Th[th:JCR]{\cite[Cor. 2.3]{Aoto}}
Let $\sim$ be an equivalence relation and
$\to$ a binary relation on the same set.
Suppose $\HH$ is a symmetric relation such that
$\HH^* \;=\; \sim$.
If 
\begin{enumerate}[noitemsep]
\item $\to$ is well-founded,
\item $\ot \circ \to \;\;\subseteq\;\; \too \circ \Heq \circ \oot$ 
(i.e., $\to$ is  \W{locally confluent in one step}), and

\item $ \ot \circ \HH \;\;\subseteq\;\; \too \circ \Heq \circ \oot$ 
(i.e., $\to$ is  \W{locally} coherent \W{in one step})
\end{enumerate}
then $\to$ is Church-Rosser modulo $\sim$.
\oTh

We apply the above theorem to
our situation.
We now take $\to$ to be $\toFOLDr$,
\bisim to be the bisimulation 
restricted on \TT, and
\HH to be a relation on \TT defined by:

\medskip

\begin{tabular}{lll}
$s \HH t$ & iff &
$\Theta \pr s = t : \vec c$ is derived from $\AxCy\union\AxBr$ 
for some \Theta and $\vec c$
\\ && by 
the cartesian second-order equational logic in Fig. \ref{fig:sel} \\ &&
without
using (Ref) and (Tra).
\end{tabular}
\medskip

\noindent
Namely,
the symmetric relation $\HH$ is the congruence closure of
one-step application of an instance of 
an axiom of $\AxCy\union\AxBr$ on \TT.
Thus $\HH^* \;=\; \bisim$.
The condition (i) of Thm. \ref{th:JCR} holds by Thm. \ref{th:fold-sn}.
We check the conditions (ii) and (iii) of Thm. \ref{th:JCR}.

\Prop[th:WCR]
The relations $\toFOLDr$  on \TT
is locally confluent in one step.
\oProp
\proof Because $\toFOLDr$ is confluent. \QED

\Prop[th:LCH]
The relation $\toFOLDr$ on \TT is locally coherent in one step.
\oProp
\proof
Let $\to$ denote $\toFOLDr$ on \TT.
We need to show that
for an instance $s=t$ of each axiom 
such that $s$ has a reduct $s'$, $s'$ and $t$ commute modulo $\sim$.
Thus we check all possible cases
of the form 
$s' \ot s \HH t$.
This is by induction on the proof of $s \HH t$.
Throughout the proof,
we write $t\sub s$ for $t\set{\vec y\mapsto \vec s}$.
Let $\vec y = y_1, \ooo, y_n,\; \vec s=s_1,\ooo,s_n$.

\begin{itemize}[leftmargin=*]
\item (Ax1)(Ax2): We check each axiom in $\AxCy\union\AxBr$.

\begin{itemize}[leftmargin=0em,label={\textbf{-}\!}]
\item (sub): 
Case $(\vec y.t) \at <\vec s> = t\sub s$.
\begin{itemize}[leftmargin=*]
\item Case $t$ is rewritten. We have
\begin{diagram}[height=1.5em]
(\vec y.t) \at <\vec s> &\rHH& t\sub s  \\
\dTo && \dTo \\
(\vec y.t') \at <\vec s> &\rHH& t'\sub s  
\end{diagram}
The case $s_i$ is rewritten  is similar.
\item If the root of $(\vec y.t) \at <\vec s>$ is rewritten,
then it is finally rewritten to $t\sub s $.
\end{itemize}

\item (sub) the converse: $t\sub s = (\vec y.t) \at <\vec s> $.\\
If $\vec s$ or $t$ is rewritten, similarly to the above case.\\
If $\vec s$ and $t$ cannot be rewritten, but
$t\sub s$ can be rewritten.
Then there exist $1\le i \le n$ and
a context $C$ such that 
$t = C[\fold(\vec z.y_i)]\quad  (y_i\not\in \vec z)$. Then $t\sub s = C[\fold(\vec z.s_i)]$,
$s_i$ is one of the patterns in \FOLDr, i.e.,
$\fold(\vec z.s_i) \to_\FOLDr r\;$, where $\vec z$ do not appear in $s_i$.

We have
\begin{diagram}[height=1.5em]
  C'[\fold(\vec z.s_i)] &\rHH& (\vec y.C[\fold(\vec z.y_i)]) \at <\vec s> \\
\dTo &&\dTo>+\\
C'[r] &\lTo& C'[\fold(\vec z.s_i)]
\end{diagram}
where $C'$ is obtained from $C$ replacing every $y_i$ in $C$ with $s_i$
for $1\le i \le n$.

\newcommand{\pionee}{(\vec y.y_1)}
\newcommand{\pitwo}{\ccc,(\vec y.y_n)}
\bigskip

\item (SP): $<\pionee\at t,\pitwo\at t> = t$.
  \begin{itemize}
  \item Case $t$ is rewritten.
\begin{diagram}[height=1.5em]
<\pionee\at t,\pitwo\at t> &\rHH& t \\
\dTo && \dTo \\
<\pionee\at t',\pitwo\at t>\\
\dTo_+ &&  \\
<\pionee\at t',\pitwo\at t'> &\rHH& t'
\end{diagram}
The converse is similar.

\item Case $t=<\vec u>$ and $(\vec y.y_i)\at t$ is rewritten.
Then $<\vec u> \ootplus <\ccc,(\vec y.y_i)\at t,\ccc> \HH <\vec u>$.
\end{itemize}

\item 
All the axioms of \AxBr, (\Bekic), (CI) and their converses:
since the roots of the left and right-had sides of each axiom are not redexes,
possible redexes appear only at the positions of metavariables
(such as $\va s,\va t$) in the axiom, hence 
these are proved similarly to (SP).

\item (dinat$_1$): $\cy{x.  s\subi{z\mapsto t}} =  s \subi{z\mapsto \cy{ z. t\subi{x\mapsto s} }}$.

\begin{enumerate}[leftmargin=*]
\item Case $s$ or $t$ is rewritten. Similarly to the case (SP).

\item Case $s$ and $t$ cannot be rewritten, and
$t\subi{x\mapsto s}$ can be rewritten.
This is only when
$s= C[\fold(\vec y.z)]$ and $t$ is 
one of the patterns in \FOLDr.
Then
\begin{itemize}
\item $s\subi t = C[\fold(\vec y.t)]$ and
\item $t\subi s = t\set{C[\fold(\vec y.x)]}$.
\end{itemize}
In this case, (dinat$_1$)'s 
rhs
has
a sub-term $\cy{ z. t \{ s\} }  = \cy{ z. s\subi{C[\fold(\vec y.z)]} }$,
which is bad. Therefore, this case is excluded.
\label{itm:ts}

\end{enumerate}

\item (dinat$_n$):
$\cy{\vec x.  s\sub{\pi_i\at t}} =  (\vec z. s) \at  \cy{\vec
  z. t\sub{\pi_i\at s} } $, where $\pi_i = \vec y.y_i$.
\begin{enumerate}[leftmargin=*]
\item Case $s$ or $t$ is rewritten. Similarly to the case (SP).

\item Case $t$ and $s$ cannot be rewritten.
If $s = <s_1,\ooo,s_n>$ and $t = <t_1,\ooo,t_n>$, reducing 
$\pi_i\at s$ and $\pi_i\at t$, similarly to \ref{itm:ts}.
Otherwise, $\cy{\vec x.  s\sub{\pi_i\at t}}$
is not reducible, hence done.
\end{enumerate}

\item (dinat$_1$)(dinat$_n$): the converses are similar.
\end{itemize}

\item (Fun) and (OSub): By induction hypothesis and 
the fact that rewriting and equational reasoning are 
closed under contexts and substitutions for variables.\qedhere
\end{itemize}

\Th
\FOLDr is \CRsim on \TT.
\oTh
\proof
The relation $\to_\FOLDr$ on \TT is SN (Thm. \ref{th:fold-sn}),
locally confluent in one step (Prop. \ref{th:WCR}),
locally coherent in one step (Prop. \ref{th:LCH}). By Thm. \ref{th:JCR},
we have \CRsim.
\QED

\Remark
There are several other criteria to establish \CRsim, which requires
termination (SN) of rewriting modulo the equivalence relation defined by
$\relmod \;\deq\; (\bisim \cdot \to \cdot \bisim)$
\cite{Huet,JouannaudKRijcai83} (see also \cite[Thm. 2.2]{Aoto}
for a unified theorem).
However, in the case of \FOLDr,
the rewriting modulo bisimulation
\relmod is not SN because (dinat) 
can copy a redex inside \fcy.
We write $\fold\;\EMP$ for $\fold(e,\EMP)$.
For example, since $\fold\;\EMP$ is a redex by (2r)
(N.B. $\cy{y.\fold\;\EMP}$ is not bad), we have an infinite derivation
\begin{meqa}
{\cy{y.\fold\;\EMP}} &\bisim \underline{(y.\fold\;\EMP)}\at \cy{y.\fold\;\EMP}\\
             &\to    (y.\EMP)     \at \underline{\cy{y.\fold\;\EMP}} \\
             &\bisim (y.\EMP)     \at \underline{(y.\fold\;\EMP)} \at \cy{y.\fold\;\EMP}\\
             &\to     (y.\EMP)  \at (y.\EMP)\at \underline{\cy{y.\fold\;\EMP}}
\; \bisim \ccc 
\end{meqa}
where each underlined term is transformed.
\oRemark

As discussed in \Sec \ref{rem:CRsim}, we have a remarkable result.

\Cor
The equational theory generated by \FOLDr, \AxCy and \AxBr
on terms of \TT is decidable.
\oCor

Since \FOLDr  is SN and \CRsim, we have the following fundamental property.

\Prop[th:bisim-fold]
If $\vec{z:b},\jud \Gamma s \vec c$ 
and $\vec{z:b},\jud \Gamma t \vec c$ are not bad, then
\begin{equation*}
s \bisim t \quad\TO\quad 
\foldE{{x_1,\ooo,x_n }. s;\; \vec x} \normalise \circ \sim\circ \normaliserev
\foldE{{x_1,\ooo,x_n }. t;\; \vec x}
\end{equation*}
where $\Gamma=x_1:b_1,\ooo,x_n:b_n$.
\oProp
\proof
If $s \bisim t$, then 
$\foldE{{x_1,\ooo,x_n }. s;\; \vec x} \bisim
 \foldE{{x_1,\ooo,x_n }. t;\; \vec x}$, hence
their normal forms are bisimilar by \CRsim.
\QED

\Remark
We introduced the notion of bad terms to establish 
the property \CRsim.
Another explanation of bad terms is that 
a bad term generates a non-rational tree.
For example, consider a function incrementing each element of 
a \code{CNat}-list defined
by
\[
  \mapinc(t) = \fold(\vnil,\; \ell.s. \ell+1\co s,\; t)
\]
This function itself is no problem. For example,
applying \code{mapinc} to
the cyclic list of 1, we have the cyclic list of 2.
\begin{Verbatim}[commandchars=\\\{\},codes=\mathcom]
  mapinc(cy(x. 1\coSym x)) \tto cy(x. mapinc(x. 1\coSym x; x)) 
  \tto cy(x. 2 \coSym mapinc(x; x)) \tto cy(x. 2\coSym x) 
\end{Verbatim}
\bigskip

\noindent
But consider a \W{bad term} $\fcy(z.1\coSym \mapinc(z))$.
Observe that it is a \W{normal form} with respect to \FOLDr,
because $\mapinc(z)$ cannot be rewritten.
But using the cartesian second-order equational logic with
\AxCy (especially (fix)) and \FOLD,
we can reason as follows:
\begin{equation}
\arraycolsep = 0.5mm
  \begin{array}[h]{llll}
\fcy(z.1\coSym \mapinc(z)) &= 1\coSym \mapinc(\fcy(z.1 \coSym \mapinc(z))) 
\\
&= 1\coSym \mapinc(1 \coSym \mapinc(\fcy(z.1\coSym \mapinc(z)))))\\
&= 1\coSym (2\coSym \mapinc(\mapinc(\fcy(z.1\coSym \mapinc(z)))))\\
&= 1\coSym 2\coSym 3\coSym \ccc \coSym n \coSym \mapinc^n(\fcy(z.1\coSym \mapinc(z)))
  \end{array}
\label{eq:mapinc}
\end{equation}
which essentially means that it corresponds to a non-rational tree.
Any cyclic term \W{without fold} is interpreted
as a rational tree in \CPO because a free iteration theory of 
binary branching trees by \AxBr
characterises rational trees modulo bisimulation 
\cite{Esik00,BE} (see also \cite[Thm. 3.2]{FICS}).
Then the continuous function \mfold in \CPO is a 
function between rational trees modulo bisimulation, 
but the example (\ref{eq:mapinc}) shows a non-rational tree.
In this sense, we excluded the bad term from \TT.

Note that a bad term $\fcy(z.1\coSym \mapinc(z))$
can be represented in Haskell as a recursive definition of list
\begin{center}
\code{z = 1:map (1+) z}  
\end{center}
which is a well-known lazy programming technique.
It looks like this is defining a cyclic data, 
but actually defining an acyclic infinite data 
described in \cite[\Sec
2.2.1]{DBLP:conf/popl/FegarasS96}, and
analysed in our previous work
\cite[\Sec 2.1]{TFPcyc}.
We could capture this phenomenon 
from a different viewpoint, namely from the viewpoint 
of rewriting. 
From the view of functional programming,
bad terms generate acyclic infinite data, and
from the view of rewriting, bad terms
are terms preventing \CRsim.
\oRemark

% Local Variables:
% TeX-master: "fs"
% End:

\section{Computing by Fold on Cyclic Datatypes}\label{sec:prog}

In this section, we demonstrate fold computation on cyclic data
by several examples. In this section, we write $\to$ to mean
$\toFOLDsimp$.

\Example[ex:emptyness]
We consider the emptiness check of cyclic sharing trees.
It means to check whether a given closed term is bisimilar 
to \nil.
This is not just to check whether a given term is syntactically \nil.
For example, is \code{cy(x.x) + cy(x.\nil\, + x)} empty?
This requires certain computation, which we will define.
First, we define a (cyclic) boolean datatype having the operation $\AND$,
satisfying the axioms \AxBr.

\bigskip
\begin{minipage}{0.4\textwidth}
\begin{Verbatim}[commandchars=\\\{\},codes=\mathcom]
ctype Bool where
  true  : Bool
  false : Bool
  $\AND$  : Bool,Bool \tto Bool
with axioms \nAxCy,\nAxBr(true,$\AND$)

\end{Verbatim}
\end{minipage}
\begin{minipage}{0.4\textwidth}
\begin{Verbatim}[commandchars=\\\{\},codes=\mathcom]
isEmpty : CTree \tto Bool
spec isEmpty(\nil)     = true
     isEmpty($f$(t))  = false
     isEmpty(s + t) = isEmpty(s) $\AND$ isEmpty(t)


\end{Verbatim}
\end{minipage}

\code{fun isEmpty(t) = fold (true, x.false, x.y.x $\AND$ y) t}
\bigskip

\noindent
We define \code{isEmpty} by \fold.
\noindent
Examples of computation are as follows, where 
the cycle cleaning laws in \SIMP are crucial.

\begin{minipage}{0.4\textwidth}
\begin{Verbatim}[commandchars=\\\{\},codes=\mathcom]
  isEmpty(cy(x.x) + cy(x.\nil\,+x)) 
\ttoo isEmpty(\nil\,+cy(x.x)) 
\tto isEmpty(\nil+\nil) 
\ttoo true $\AND$ true \tto true



\end{Verbatim}
\end{minipage}
\quad
\begin{minipage}{0.4\textwidth}
\begin{Verbatim}[commandchars=\\\{\},codes=\mathcom]
isEmpty( cy(x$^{\code{Ctree}}$.cy(w.x)) )
\tto cy(x$^{\code{Bool}}$. isEmpty(x.cy(w.x); x))
\tto cy(x$^{\code{Bool}}$. cy( w.isEmpty(x.w.x; x,w) ))
\tto cy(x$^{\code{Bool}}$. cy( w.x )) 
\tto cy(x$^{\code{Bool}}$.x) \tto true

isEmpty(cy(x. a(cy(y.y+y)) + cy(w.x))) 
\ttoo isEmpty(cy(x. a(\nil) + x)) 
\tto isEmpty(a(\nil)) \tto false
\end{Verbatim}
\end{minipage}
\bigskip

The example \code{isEmpty(cy(x.x) + cy(x.\nil\,+x))}
is essentially checking null-ability of 
a grammar \cite{nullgrammer}, which 
was pointed 
as a difficult problem 
for computing with cyclic data \cite{GrICFP}.
Our rewrite system \FOLDr with \SIMP successfully computes it 
in a principled manner (i.e., without any special treatment).
\oExample

\Example
As an example of primitive recursion on cyclic datatypes mentioned 
in \Sec \ref{sec:prim},
we consider the tail of a cyclic list, which
we call \code{ctail}. 
It should satisfy the specification
below right.
But how to define the tail of a \code{cy}-term is not immediately clear.
For example, what should be the result of 
\code{ctail (cy(x.1$\coSym$2$\coSym$x))}?
This case may need %(complicated) 
unfolding of 
cycle as in \cite{TFPcyc}.
A naive unfolding by using the fixed point law 
$\cy{x.t} = t\,\subi{x\mapsto \cy{x.t}}$
violates strong normalisation
because it copies the original term.
It actually \W{increases} complexity.

\begin{CVerbatim}[commandchars=\\\{\},codes=\mathcom]
ctail : CList $\to$ CList
spec ctail (\nil)       = \nil
     ctail (k \con t)     = t
     ctail (cy(x. t))= ??
\end{CVerbatim}

We define \code{ctail} by \fold.
Rather than the fixed point law, 
another important principle of cyclic structures of \Hi{\Bekic law}
plays a crucial role here.

\begin{CVerbatim}[commandchars=\\\{\},codes=\mathcom]
fun ctail(t) = \pione fold (<[],[]>, k,x,y.<y, k \con y>) t

ctail(cy(x.1 \con 2 \con x)) \ttoo \pione cy(x,y. <2 \con y, 1 \con 2 \con y>) 
    \colorbox{mycolor}{\parbox[t]{.5cm}{$\to^+$}} \pione <cy(x.2 \con cy(y.1 \con 2 \con y)), cy(y.1 \con 2 \con y)>
     \tto  cy(x.2 \con cy(y.1 \con 2 \con y)) \tto 2 \con cy(y.1 \con 2 \con y)     \RM{\textbf{(Normal form)}}
\end{CVerbatim}
\noindent
Note that the above normal form does not mean a head normal form and
we do not rely on lazy evaluation.
The highlighted step uses \Bekic law.
\oExample

\Example
We define the datatype of cyclic strings
consisting of \code{a}, \code{b}, the empty string \epsilon, and
the choice operator ``\code{|}''.

\medskip
\begin{Verbatim}[commandchars=\\\{\},codes=\mathcom]
  ctype CString where
    a : CString \tto CString
    b : CString \tto CString
    \epsilon : CString
    |  : CString,CString \tto CString
  with axioms \nAxCy,\nAxBr($\epsilon$,|)
\end{Verbatim}
\medskip

\noindent
We consider the function \fnaa that checks whether 
a given closed term contains two consecutive \code{a}'s, such as
\code{a(a($\cdot$))}.
For example, 

\medskip
\begin{Verbatim}[commandchars=\\\{\},codes=\mathcom]
   aa?( b(a(a(b(\epsilon)))) ) $\quad\normalise\quad$ true
   aa?( b(b(\epsilon))         $\quad\normalise\quad$ false
   aa?( b(\epsilon)|a(a(\epsilon)) )  $\quad\normalise\quad$ true
\end{Verbatim}
\medskip

\noindent
Even in case of a string containing cycle, the function must 
return a correct result.
For example, we expect the following results.

\medskip
\begin{Verbatim}[commandchars=\\\{\},codes=\mathcom]
   aa?( cy(x.a(x)) )          $\quad\normalise\quad$ true
   aa?( b(cy(x.a(b(a(x))))) ) $\quad\normalise\quad$ true
\end{Verbatim}
\medskip

\noindent
\noindent
We now specify \code{aa?} by using the function \code{head-a?}
which checks whether the head is \code{a}.

\medskip
\begin{minipage}{0.4\textwidth}
\begin{Verbatim}[commandchars=\\\{\},codes=\mathcom]
head-a? : CString \tto ABool
spec head-a?(a(t)) = true
     head-a?(b(t)) = false


\end{Verbatim}
\end{minipage}
\qquad\qquad
\begin{minipage}{0.4\textwidth}
\begin{Verbatim}[commandchars=\\\{\},codes=\mathcom]
aa? : CString \tto ABool
spec aa?(a(t)) = head-a?(t) 
     aa?(b(t)) = aa?(t)
     aa?( \epsilon ) = false
     aa?(s | t) = aa?(s) $\vee$ aa?(t)
\end{Verbatim}
\end{minipage}
\bigskip

\noindent
The results of \code{aa?} must be boolean, but it should not be the
type \code{Bool} defined in Example \ref{ex:emptyness}.
Now we need the type \code{ABool} of ``additive'' boolean, meaning
that it has the conjunction ``$\OR$'' with the unit \code{false},
because the specification of \code{aa?} requires that
the unit $\epsilon$ (resp. the multiplication ``\code{|}'') of \code{CString} 
is mapped to the unit \code{false} (resp. the multiplication ``$\OR$'') 
of the target type.

\begin{CVerbatim}[commandchars=\\\{\},codes=\mathcom]
  ctype ABool where
    true  : ABool
    false : ABool
    $\OR$     : ABool,ABool \tto ABool
  with axioms \nAxCy,\nAxBr(false,$\OR$)
\end{CVerbatim}

\noindent
Then we can define the functions \code{head-a?} and \code{aa?} by fold.

\begin{CVerbatim}[commandchars=\\\{\},codes=\mathcom]
  fun  head-a?(t) = fold (x.true,x.false,false,x.y.x$\OR$y) t
  fun  aa?(t) = \pione fold (v,w.<head-a?(w),a(w)>, v,w.<v,b(w)>
                          <false,$\epsilon$>, v$_1$.w$_1$.v$_2$.w$_2$.<v$_1\OR$v$_2$,w$_1$|w$_2$>) t
\end{CVerbatim}

\noindent
We demonstrate how our system correctly computes 
\begin{center}
\code{aa?(cy(x.a(x)))} $\quad\normalise\quad$ \code{true}.
\end{center}
One may think that it needs expansion of 
the inner term of \code{cy}.
We show that the rewrite system \FOLDsimp does it \W{without using the fixed point law},
rather, by using 
\W{\Bekic law}.
We write \code{folde} %is short 
for
\code{fold (v.w.<head-a?(w),a(w)>, v.w.<v,b(w)>,\ooo)}.

\begin{CVerbatim}[commandchars=\\\{\},codes=\mathcom]
aa?(cy(x.a(x))) = \pione folde cy(x.a(x))
 \tto \pione cy(v.w.(folde a(v); v,w))
\colorbox{mycolor}{\parbox[t]{.5cm}{$\to^+$}} \pione <cy(v.head-a?(cy(w.a(w)))), cy(w.a(w))>
 \ttoo head-a?(cy(w.a(w)) \tto cy(w.head-a?(w.a(w);w) ) \ttoo cy(w.true) \tto true
\end{CVerbatim}
Again, the highlighted step uses \Bekic law.
Moreover, we expect 
\begin{center}
\code{aa?(cy(x.b(x)))} $\quad\normalise\quad$ \code{false}
\end{center}
without falling into non-termination. This is obtained
without any special treatment.

\begin{CVerbatim}[commandchars=\\\{\},codes=\mathcom]
aa?(cy(x.b(x))) = \pione folde cy(x.b(x))
 \tto \pione cy(v.w. (folde b(v); v,w))
\colorbox{mycolor}{\parbox[t]{.5cm}{$\to^+$}} \pione <cy(v.v),cy(w.b(w))> \ttoo cy($\code{v}$.v) \tto false
\end{CVerbatim}

\noindent
A crucial point is that since we chose the target type as \code{ABool},
the rule (14r) in \FOLDsimp rewrites
\code{cy($\code{v}$.v)} to \code{false}, which is 
the unit of \code{ABool}.
If we chose the target type as \code{Bool} in Example \ref{ex:emptyness},
this would become an incorrect result \code{true}.
But the specification forced us to choose \code{ABool}.
\oExample

\Example
This example shows that our cyclic datatype has ability to express
graphs.
The graph shown below right represents friend relationship,
which describes Bob knows Alice, and Carol knows Alice and Bob.
To make it a rooted graph, it has the uppermost node ``+'' which points to
the nodes of three persons.
This is represented as a term

\medskip
\begin{Verbatim}[commandchars=\\\{\},codes=\mathcom]
  (a.b.c.a+b+c) $\at$ cy(a.b.c.<name("alice"), name("bob")+knows(a), 
                            name("carol")+knows(a)+knows(b)>)
\end{Verbatim}
\medskip

\noindent
which we call \code g. The term \code g 
is of type \code{FriendGraph}
defined as follows.

\medskip
\begin{minipage}{0.4\textwidth}
\begin{Verbatim}[commandchars=\\\{\},codes=\mathcom]
ctype FriendGraph where
  knows : FriendGraph \tto FriendGraph
  name  : String \tto FriendGraph
  \nil : FriendGraph
  + : FriendGraph,FriendGraph \tto FriendGraph
with axioms \nAxCy,\nAxBr(\nil,+)
\end{Verbatim}
\end{minipage}
\x{5em}
\begin{minipage}{0.4\textwidth}
  \begin{center}
\includegraphics[scale=.4]{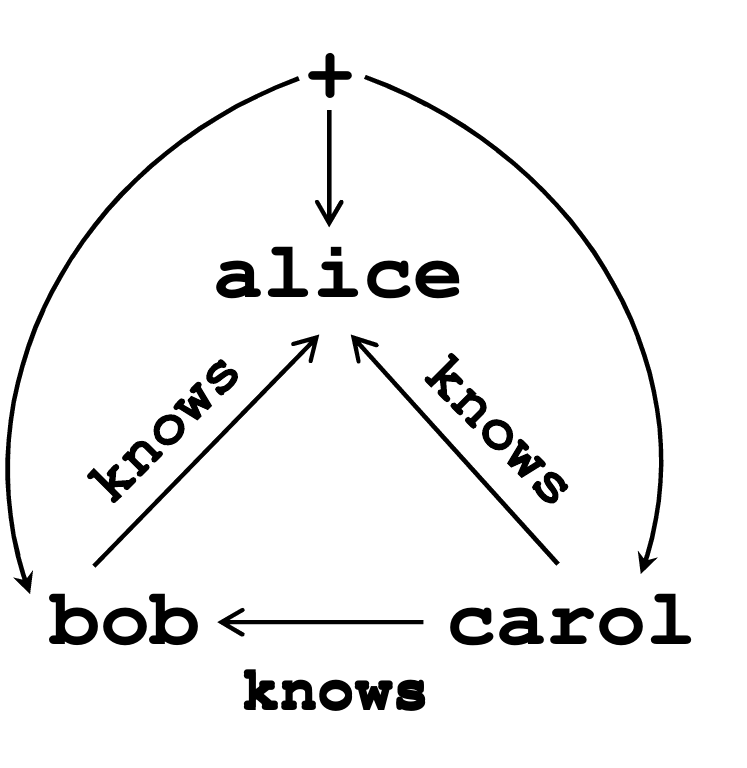}    
  \end{center}
\end{minipage}
                                
\noindent
We define a function \code{collect}
that collects all names in a graph as a name list
of type \code{Names}.

\bigskip
\x{-2em}
\begin{minipage}{0.4\textwidth}
\begin{Verbatim}[commandchars=\\\{\},codes=\mathcom]
ctype Names where
  nm : String \tto Names 
  \nil : Names
  + : Names,Names \tto Names
with axioms \nAxCy,\nAxBr(\nil,+)

\end{Verbatim}
\end{minipage}
\quad
\begin{minipage}{0.4\textwidth}
\begin{Verbatim}[commandchars=\\\{\},codes=\mathcom]
collect : FriendGraph \tto Names
spec collect (knows(t)) = collect(t)
     collect (name(p)) = nm(p)


\end{Verbatim}
\end{minipage}
\begin{Verbatim}[commandchars=\\\{\},codes=\mathcom]
fun collect t = \pione folde (x,y.<x,knows(y)>, x,y.<nm(y),name(y)>,
                           <\nil,\nil>, v$_1$.w$_1$.v$_2$.w$_2$.<v$_1+$v$_2$,w$_1+$w$_2$>) t
\end{Verbatim}

\bigskip

\noindent
Then we collect certainly all names by our system as follows,
where \code{folde} is short for \\
\code{fold (x,y.<x,knows(y)>, x,y.<nm(y),name(y)>,\ooo)}.

\medskip
\begin{Verbatim}[commandchars=\\\{\},codes=\mathcom,fontsize=\small]
collect g = \pione folde g 
\tto \pione (a.a'.b.b'.c.c'.<a+b+c,a'+b'+c'>) $\at$ (cy(a.a'.b.b'.c.c'.
    <folde(a.b.c.name("alice")), folde(a.b.c.name("bob")+knows(a)), 
     folde(a.b.c.name("carol")+knows(a)+knows(b)>)))
\ttoo \pione (a.a'.b.b'.c.c'.<a+b+c,a'+b'+c'>) $\at$ (cy(a.a'.b.b'.c.c'.
  < <nm("alice"),name("alice")>, <nm("bob"),name("bob")>, <nm("carol"),name("carol")> >)
\ttoo nm("alice")+nm("bob")+nm("carol")
\end{Verbatim}

\oExample

% Local Variables:
% TeX-master: "fs"
% End:

\section{Related work}

\subsec{Cyclic structures represented by systems of equations}
Other than the term representation used in the present paper,
various representations of cyclic structures or graphs are known.
One of the frequently used representations is by
a \Hi{system of equations}, also known as
an {equational term graph} \cite{ETGRS}.
It is essentially the same as the representation of cyclic structures using
\textbf{letrec}-expressions used in \cite{Hassei}.
Semantically, the least solution of a system of equations is
regarded as a (unfolded) graph.
Syntactically, a system of (flat)
equations can be seen as adjacency lists of a graph, e.g.
an equation $x = f(y_1,\ooo,y_n)$ in a system can be regarded as an 
adjacency list of vertex $x$ pointed to by
other vertices $y_1,\ooo,y_n$ 
(cf. the discussions in \cite[Sect. 8]{LMCS}).
These representations and our term representation are equivalent. For example,
a system of equations
\[
\left\{
  \begin{array}{lll}
x_1 = f(x_1,x_2),\\
x_2 = g(x_1)
  \end{array}
\right\}
\]
is equivalently represented as a term
\[
\cy{x_1,x_2. <f(x_1,x_2),g(x_1)>}.
\]

Nishimura and Ohori \cite{Nishimura,NiOh99}
developed a general mechanism for programming with 
cyclic structures in a purely functional style using the representation
of system of equations.
They developed the ``reduce'' operation on cyclic structures
based on a mechanism
for data-parallelism on recursive data, which is 
\fold in our sense,
although bisimilar cyclic structures are not identified in their work.

\subsection{Foundational graph rewriting calculi}
There has been various work to
deal with graph computation and 
cyclic data structures %nicely
in functional programming and foundational calculi
including \cite{DBLP:conf/popl/FegarasS96,TFPcyc,tlca,LMCS,FLOPS,GrICFP,
ICFP10,MA,ETGRS,cyclic-lmd}.
Foundational graph rewriting calculi, such as equational term graph
rewriting systems \cite{ETGRS}, are general frameworks of graph computation.
The \fold on a cyclic datatype in this paper is 
more restricted than general graph
rewriting. However, our emphasis is 
clarification of the categorical and algebraic structures of 
cyclic datatypes and the computation by \fold on them 
by regarding \fold as a structure preserving map, rather than
unrestricted rewriting.
This was a key to obtain SN and \CRsim of \FOLDr.
We also hope that it will be useful for further optimisation 
such as the fold fusion based on semantics  as done in \cite[\Sec 4.3]{MSCS}.
The general study of graph rewriting was also important for our
study at the foundational level.
The unit ``$\vnil$'' of branching in \AxBr
corresponds to
the black hole constant ``\oo'' considered in \cite{ETGRS}, due to \cite{BE}.
This observation has been used to give an effective operational semantics of graph
transformation in \cite{MA}.

\section{Conclusion}
\label{sec:conc}

In this paper, we have developped foundations of cyclic datatypes
and computation:

\def\labelenumi  {\theenumi}
\def\theenumi    {[\Roman{enumi}]}
\begin{enumerate}%[nosep]
\item Syntax and type system supporting algebraic datatypes with cycle and 
sharing constructs (\Sec \ref{sec:salg})
\item Complete equational axioms for bisimulation of 
cyclic data (Fig. \ref{fig:axioms})
\item Algebraic theory \FOLD of fold on cyclic datatypes
and its strongly normalising rewrite system \FOLDr (Fig. \ref{fig:FOLDr},
Fig. \ref{fig:fold}) (\Sec \ref{sec:fold}, \Sec \ref{sec:SN}),
which is Church-Rosser modulo bisimulation (\Sec \ref{sec:CR})
\item The framework that supports [I]-[III] based on 
cartesian second-order algebraic theory (\Sec \ref{sec:soa})
and iteration category (\Sec \ref{sec:cat}).
The numbered items [I]-[III] in Fig. \ref{fig:clist-view} in Introduction
are instances of these results.
\end{enumerate}

\noindent We have not assumed any particular operational semantics nor
strategy to obtain strongly normalising
\fold on cyclic data. This point may be useful to
deal with cyclic datatypes in proof assistances
requiring terminating functions, such as Coq or Agda.
We have shown several concrete examples of cyclic data computation in \Sec
\ref{sec:prog}. 
In this paper, we have focused on the underlying theory 
of cyclic datatypes.
Formal development of the programming language that realises
the program codes described in this paper is left for a future work.

\subsec{Acknowledgments} 
I am grateful to  Kazutaka Matsuda and Kazuyuki Asada
for various discussions about calculi and programming languages 
about graphs and cyclic structures, and
Fr\'{e}d\'{e}ric Blanqui
for clarifying some details of the General Schema.
I also thank the reviewers for their positive and constructive comments. 

% Local Variables:
% TeX-master: "fs"
% End:

{
\bibliographystyle{alpha}
\bibliography{bib}

\newcommand{\etalchar}[1]{$^{#1}$}
\begin{thebibliography}{GHUV06}

\bibitem[AB97]{cyclic-lmd}
Z.~M. Ariola and S.~Blom.
\newblock Cyclic lambda calculi.
\newblock In {\em Theoretical Aspects of Computer Software, LNCS 1281}, pages
  77--106, 1997.

\bibitem[Acz78]{Aczel}
P.~Aczel.
\newblock A general {Church-Rosser} theorem.
\newblock Technical report, University of Manchester, 1978.

\bibitem[AJ94]{Abramsky-Jung}
S.~Abramsky and A.~Jung.
\newblock Domain theory.
\newblock In D.~Gabbay and T.~S.~E. Maibaum, editors, {\em Handbook of Logic in
  Computer Science}, volume~3, pages 1--168. Oxford University Press, 1994.

\bibitem[AK96]{ETGRS}
Z.~M. Ariola and J.~W. Klop.
\newblock Equational term graph rewriting.
\newblock {\em Fundam. Inform.}, 26(3/4):207--240, 1996.

\bibitem[AT12]{Aoto}
T.~Aoto and Y.~Toyama.
\newblock A reduction-preserving completion for proving confluence of
  non-terminating term rewriting systems.
\newblock {\em Logical Methods in Computer Science}, 8(1), 2012.

\bibitem[B{\'{E}}93]{BE}
S.~L. Bloom and Z.~{\'{E}}sik.
\newblock {\em Iteration Theories -- The Equational Logic of Iterative
  Processes}.
\newblock {EATCS} Monographs on Theoretical Computer Science. Springer, 1993.

\bibitem[BFS00]{Buneman}
P.~Buneman, M.~F. Fernandez, and D.~Suciu.
\newblock {UnQL}: {A} query language and algebra for semistructured data based
  on structural recursion.
\newblock {\em {VLDB} J.}, 9(1):76--110, 2000.

\bibitem[BJO02]{IDTS}
F.~Blanqui, J.-P. Jouannaud, and M.~Okada.
\newblock Inductive data type systems.
\newblock {\em Theoretical Computer Science}, 272:41--68, 2002.

\bibitem[Bla00]{IDTS00}
F.~Blanqui.
\newblock Termination and confluence of higher-order rewrite systems.
\newblock In {\em Rewriting Techniques and Application (RTA 2000)}, LNCS 1833,
  pages 47--61. Springer, 2000.

\bibitem[Bla16]{Blanqui-TCS}
F.~Blanqui.
\newblock Termination of rewrite relations on \lmd-terms based on {Girard}'s
  notion of reducibility.
\newblock {\em Theor. Comput. Sci.}, 611:50--86, 2016.

\bibitem[BN98]{Baader}
F.~Baader and T.~Nipkow.
\newblock {\em Term Rewriting and All That}.
\newblock Cambridge University Press, 1998.

\bibitem[Brz64]{nullgrammer}
J.~A. Brzozowski.
\newblock Derivatives of regular expressions.
\newblock {\em J. ACM}, 11, 1964.

\bibitem[CGZ05]{contextlogic}
C.~Calcagno, P.~Gardner, and U.~Zarfaty.
\newblock Context logic and tree update.
\newblock In {\em Proc. of POPL'05}, pages 271--282, 2005.

\bibitem[Chl08]{PhoasICFP08}
A.~Chlipala.
\newblock Parametric higher-order abstract syntax for mechanized semantics.
\newblock In {\em ICFP'08: Proceedings of the 13th ACM SIGPLAN International
  Conference on Functional Programming}, 2008.

\bibitem[Coq92]{CoqPat}
T.~Coquand.
\newblock Pattern matching with dependent types.
\newblock In {\em Proc. of the 3rd Work. on Types for Proofs and Programs},
  1992.

\bibitem[Cou83]{InfTree}
B.~Courcelle.
\newblock Fundamental properties of infinite trees.
\newblock {\em Theoretical Computer Science}, 25(2):95--169, 1983.

\bibitem[DFH95]{HOAS-Coq}
J.~Despeyroux, A.~Felty, and A.~Hirschowitz.
\newblock Higher-order abstract syntax in {Coq}.
\newblock In {\em Typed Lambda Calculi and Applications, LNCS 902}, pages
  124--138, 1995.

\bibitem[DPP04]{eff-bisim}
A.~Dovier, C.~Piazza, and A.~Policriti.
\newblock An efficient algorithm for computing bisimulation equivalence.
\newblock {\em Theoretical Computer Science}, 311:221--256, 2004.
\newblock Issues 1-3.

\bibitem[{\'{E}}si99]{IterCat}
Z.~{\'{E}}sik.
\newblock Axiomatizing iteration categories.
\newblock {\em Acta Cybernetica}, 14:65--82, 1999.

\bibitem[{\'{E}}si00]{Esik00}
Z.~{\'{E}}sik.
\newblock Axiomatizing the least fixed point operation and binary supremum.
\newblock In {\em Proc. of Computer Science Logic 2000}, LNCS 1862, pages
  302--316, 2000.

\bibitem[FC13]{FiorePROP}
M.~P. Fiore and M.~D. Campos.
\newblock The algebra of directed acyclic graphs.
\newblock In {\em Computation, Logic, Games, and Quantum Foundations}, LNCS
  7860, pages 37--51, 2013.

\bibitem[FH10]{2ndCSL}
M.~Fiore and C.-K. Hur.
\newblock Second-order equational logic.
\newblock In {\em Proc. of CSL'10}, LNCS 6247, pages 320--335, 2010.

\bibitem[Fio08]{Fiore2nd}
M.~Fiore.
\newblock Second-order and dependently-sorted abstract syntax.
\newblock In {\em Proc. of LICS'08}, pages 57--68, 2008.

\bibitem[FM10]{2ndAlg}
M.~Fiore and O.~Mahmoud.
\newblock Second-order algebraic theories.
\newblock In {\em Proc. of MFCS'10}, LNCS 6281, pages 368--380, 2010.

\bibitem[FPT99]{FPT}
M.~Fiore, G.~Plotkin, and D.~Turi.
\newblock Abstract syntax and variable binding.
\newblock In {\em Proc. of LICS'99}, pages 193--202, 1999.

\bibitem[FS96]{DBLP:conf/popl/FegarasS96}
L.~Fegaras and T.~Sheard.
\newblock Revisiting catamorphisms over datatypes with embedded functions (or,
  programs from outer space).
\newblock In {\em POPL'96}, pages 284--294, 1996.

\bibitem[FS14]{SamLICS}
M.~Fiore and S.~Staton.
\newblock Substitution, jumps, and algebraic effects.
\newblock In {\em Proc. of LICS'14}, 2014.

\bibitem[GHUV06]{TFPcyc}
N.~Ghani, M.~Hamana, T.~Uustalu, and V.~Vene.
\newblock Representing cyclic structures as nested datatypes.
\newblock In {\em Proceedings of Trends in Functional Programming}, pages
  173--188, 2006.

\bibitem[Ham04]{free}
M.~Hamana.
\newblock Free {$\Sigma$}-monoids: A higher-order syntax with metavariables.
\newblock In {\em Proc. of APLAS'04}, LNCS 3302, pages 348--363, 2004.

\bibitem[Ham05]{CRS}
M.~Hamana.
\newblock Universal algebra for termination of higher-order rewriting.
\newblock In {\em Proc. of RTA'05}, LNCS 3467, pages 135--149, 2005.

\bibitem[Ham07]{HSL}
M.~Hamana.
\newblock Higher-order semantic labelling for inductive datatype systems.
\newblock In {\em Proc. of PPDP'07}, pages 97--108. ACM Press, 2007.

\bibitem[Ham09]{tlca}
M.~Hamana.
\newblock Initial algebra semantics for cyclic sharing structures.
\newblock In {\em Proc. of TLCA'09}, LNCS 5608, pages 127--141, 2009.

\bibitem[Ham10]{LMCS}
M.~Hamana.
\newblock Initial algebra semantics for cyclic sharing tree structures.
\newblock {\em Logical Methods in Computer Science}, 6(3), 2010.

\bibitem[Ham12]{FLOPS}
M.~Hamana.
\newblock Correct looping arrows from cyclic terms: Traced categorical
  interpretation in {Haskell}.
\newblock In {\em Proc. of FLOPS'12}, LNCS 7294, pages 136--150, 2012.

\bibitem[Ham15]{FICS}
M.~Hamana.
\newblock Iteration algebras for {UnQL} graphs and completeness for
  bisimulation.
\newblock In {\em Proc. of Fixed Points in Computer Science (FICS'15)},
  Electronic Proceedings in Theoretical Computer Science 191, pages 75--89,
  2015.

\bibitem[Ham16]{fscd-cyc}
M.~Hamana.
\newblock Strongly normalising cyclic data computation by iteration categories
  of second-order algebraic theories.
\newblock In {\em Proc. of The 1st International Conference on Formal
  Structures for Computation and Deduction (FSCD'16)}, volume~52 of {\em the
  Leibniz International Proceedings in Informatics (LIPIcs)}, pages
  21:1--21:18, 2016.

\bibitem[Has97]{Hassei}
M.~Hasegawa.
\newblock {\em Models of Sharing Graphs: A Categorical Semantics of
  \textsf{let} and \textsf{letrec}}.
\newblock PhD thesis, University of Edinburgh, 1997.

\bibitem[HHI{\etalchar{+}}10]{ICFP10}
S.~Hidaka, Z.~Hu, K.~Inaba, H.~Kato, K.~Matsuda, and K.~Nakano.
\newblock Bidirectionalizing graph transformations.
\newblock In {\em Proc. of {ICFP} 2010}, pages 205--216, 2010.

\bibitem[HMA]{MSCS}
M.~Hamana, K.~Matsuda, and K.~Asada.
\newblock The algebra of recursive graph transformation language {UnCAL}:
  Complete axiomatisation and iteration categorical semantics.
\newblock to appear in Mathematical Structures in Computer Science, Cambridge
  University Press.

\bibitem[Hue80]{Huet}
G.~Huet.
\newblock Confluent reductions: Abstract properties and applications to term
  rewriting systems.
\newblock {\em Journal of ACM}, 27(4):797--821, 1980.

\bibitem[JKR83]{JouannaudKRijcai83}
J.~Jouannaud, H.~Kirchner, and J.~Remy.
\newblock Church-rosser properties of weakly terminating term rewriting
  systems.
\newblock In {\em Proceedings of the 8th International Joint Conference on
  Artificial Intelligence. Karlsruhe, FRG, August 1983}, pages 909--915, 1983.

\bibitem[JSV96]{JSV}
A.~Joyal, R.~Street, and D.~Verity.
\newblock Traced monoidal categories.
\newblock {\em Mathematical Proceedings of the Cambridge Philosophical
  Society}, 119(3):447--468, 1996.

\bibitem[Klo80]{Klop}
J.W. Klop.
\newblock {\em Combinatory Reduction Systems}.
\newblock PhD thesis, CWI, Amsterdam, 1980.
\newblock volume 127 of Mathematical Centre Tracts.

\bibitem[MA15]{MA}
K.~Matsuda and K.~Asada.
\newblock Graph transformation as graph reduction: A functional reformulation
  of graph-transformation language {UnCAL}.
\newblock Technical Report GRACE-TR 2015-01, National Institute of Informatics,
  January 2015.

\bibitem[Mee92]{para}
L.~G. L.~T. Meertens.
\newblock Paramorphisms.
\newblock {\em Formal Asp. Comput.}, 4(5):413--424, 1992.

\bibitem[MFP91]{DBLP:conf/fpca/MeijerFP91}
E.~Meijer, M.~M. Fokkinga, and R.~Paterson.
\newblock Functional programming with bananas, lenses, envelopes and barbed
  wire.
\newblock In {\em Proc of FPCA'91}, pages 124--144, 1991.

\bibitem[Mil84]{MilnerRegular}
R.~Milner.
\newblock A complete inference system for a class of regular behaviours.
\newblock {\em J. Comput. Syst. Sci.}, 28(3):439--466, 1984.

\bibitem[Nis97]{Nishimura}
S.~Nishimura.
\newblock A strict functional language with cyclic recursive data.
\newblock {\em Formal Asp. Comput.}, 9(1):78--97, 1997.

\bibitem[NO99]{NiOh99}
S.~Nishimura and A.~Ohori.
\newblock Parallel functional programming on recursively defined data via
  data-parallel recursion.
\newblock {\em J. Funct. Program.}, 9(4):427--462, 1999.

\bibitem[OC12]{GrICFP}
B.~C.d.S. Oliveira and W.~R. Cook.
\newblock Functional programming with structured graphs.
\newblock In {\em Proc. of ICFP'12}, pages 77--88, 2012.

\bibitem[SP82]{Smyth-Plotkin}
M.~B. Smyth and G.~D. Plotkin.
\newblock The category-theoretic solution of recursive domain equations.
\newblock {\em SIAM J. Comput}, 11(4):763--783, 1982.

\bibitem[SP00]{alex-plot}
A.~K. Simpson and G.~D. Plotkin.
\newblock Complete axioms for categorical fixed-point operators.
\newblock In {\em Proc. of LICS'00}, pages 30--41, 2000.

\bibitem[Sta13]{SamFOS}
S.~Staton.
\newblock An algebraic presentation of predicate logic.
\newblock In {\em Proc. of {FOSSACS} 201}, pages 401--417, 2013.

\bibitem[Sta15]{SamQ}
S.~Staton.
\newblock Algebraic effects, linearity, and quantum programming languages.
\newblock In {\em Proc. of POPL'15}, pages 395--406, 2015.

\bibitem[Ter03]{Terese}
Terese.
\newblock {\em Term Rewriting Systems}.
\newblock Number~55 in Cambridge Tracts in Theoretical Computer Science.
  Cambridge University Press, 2003.

\bibitem[Win93]{Winskel}
G.~Winskel.
\newblock {\em The Formal Semantics of Programming Languages}.
\newblock The MIT Press, 1993.

\end{thebibliography}
}

\end{document}